\PassOptionsToPackage{hyphens}{url}
\documentclass[sigconf,screen]{acmart}

\acmConference[ISCA 2025]{The 52nd IEEE/ACM International Symposium on Computer
Architecture}{June 21--25,
  2025}{Tokyo, Japan}



\usepackage{algorithmic}
\usepackage{textcomp}
\usepackage{fancyhdr}
\usepackage{microtype}[final]
\usepackage{hyperref}
\usepackage[linesnumbered,ruled,vlined]{algorithm2e}
\usepackage[labelformat=simple]{subcaption}

\definecolor{LightGray}{gray}{0.95}
\usepackage{multirow}
\usepackage{colortbl}
\usepackage[bottom]{footmisc}
\usepackage{dblfloatfix}
\usepackage{soulutf8}
\usepackage{enumitem}
\usepackage{listings}

\usepackage{mathtools}
\newcommand{\subheading}[1]{\noindent \textbf{#1:}}

\newcommand{\rev}[1]{\textcolor{black}{#1}}

\DeclarePairedDelimiter{\ceil}{\lceil}{\rceil}

\copyrightyear{2025}
\acmYear{2025}
\setcopyright{cc}
\setcctype{by-nc-nd}
\acmConference[ISCA '25]{Proceedings of the 52nd Annual International
Symposium on Computer Architecture}{June 21--25, 2025}{Tokyo, Japan}
\acmBooktitle{Proceedings of the 52nd Annual International Symposium on
Computer Architecture (ISCA '25), June 21--25, 2025, Tokyo,
Japan}\acmDOI{10.1145/3695053.3730991}
\acmISBN{979-8-4007-1261-6/2025/06}
\thanks{Corresponding author: smaurya@wisc.edu}

\usepackage{etoolbox}

\makeatletter
\g@addto@macro\UrlBreaks{
  \do\/
  \do-
}
\makeatother

\begin{document}

% \newcommand{\iscasubmissionnumber}{37}

% \newpage

% \input{inputs/cover}

% \newpage

\title{Synchronization for Fault-Tolerant Quantum Computers}
% \subtitle{\normalsize{ISCA 2025 Submission
%     \textbf{\#\iscasubmissionnumber} -- Confidential Draft -- Do NOT Distribute!!}}

%\author{...} % removed for anonymity
\author{Satvik Maurya}
\affiliation{
  \institution{University of Wisconsin-Madison}
  \city{Madison, WI}
  \country{USA}
}
\author{Swamit Tannu}
\affiliation{
  \institution{University of Wisconsin-Madison}
  \city{Madison, WI}
  \country{USA}
}

\begin{abstract}
Quantum Error Correction (QEC) codes store information reliably in logical qubits by encoding them in a larger number of less reliable qubits. 
The surface code, known for its high resilience to physical errors, is a leading candidate for fault-tolerant quantum computing (FTQC). 
Logical qubits encoded with the surface code can be in different phases of their syndrome generation cycle, thereby introducing desynchronization in the system. This can occur due to the production of non-Clifford states, dropouts due to fabrication defects, and the use of other QEC codes with the surface code to reduce resource requirements. 
Logical operations require the syndrome generation cycles of the logical qubits involved to be synchronized. This requires the leading qubit to pause or slow down its cycle, allowing more errors to accumulate before the next cycle, thereby increasing the risk of uncorrectable errors.

To synchronize the syndrome generation cycles of logical qubits, we define three policies - \textit{Passive}, \textit{Active}, and \textit{Hybrid}. The \textit{Passive} policy is the baseline, and the simplest, wherein the leading logical qubits idle until they are synchronized with the remaining logical qubits. On the other hand, the \textit{Active} policy aims to slow the leading logical qubits down gradually, by inserting short idle periods before multiple code cycles. This approach reduces the logical error rate (LER) by up to 2.4$\times$ compared to the \textit{Passive} policy. The \textit{Hybrid} policy further reduces the LER by up to 3.4$\times$ by reducing the synchronization slack and running a few additional rounds of error correction.
Furthermore, the reduction in the logical error rate with the proposed synchronization policies enables a speedup in decoding latency of up to 2.2$\times$ with a \textit{circuit-level} noise model.

\end{abstract}

\maketitle % should come after the abstract

\section{Introduction}

% \begin{figure}[t]
%     \centering
%     \includegraphics[width=0.5\linewidth]{figs/intro_clk_domains.pdf}
%     \caption{}
%     \label{fig:intro_clk_domains}
%     \Description[Introductory figure]{}
% \end{figure}

\begin{figure*}[t!]
    \centering
    % \includegraphics[width=0.95\textwidth]{figs/fig1_new.pdf}
    % % 
    % \caption{(a) An ideal system with uniform latencies for all operations -- all patches will be in concert ensuring easy implementation of Lattice Surgery for performing non-Clifford operations; (b) Impact of non-uniform latencies on the logical qubits -- $P'$ is lagging behind $P$, thus introducing a slack between the two logical qubits that must be absorbed before performing Lattice Surgery; (c) An expanded view of the stabilizer circuits of the two skewed patches (only one plaquette shown) in (b) -- $P$ at time $\mathbf{t}$ is executing the final layer of Hadamard gates while $P'$ at time $\mathbf{t}$ is on the first CNOT layer; (d) Patch $P$ starts round $r$ before patch $P'$ -- in \textit{Active} synchronization, the slack is distributed evenly between the $d$ rounds while in \textit{Passive} synchronization, the slack is added at the end of the last round before synchronization and Lattice Surgery. 
    % % {\color{red} Swamit --  change ancillary qubits to parity in fig (c), show d rounds on merged patch fig (d), In (a) in the gray box add LRC gadgets}
    % % (a) Disparity between expected lockstep execution of QEC with uniform gate latency and a realistic logical qubit with non-uniform gate latency; (b) Runahead execution to minimize qubit idle time; (c) Synchronization primitives for logical operation using Lattice Surgery.
    % }
    \hspace*{\fill}
    \begin{subfigure}[b]{0.22\linewidth}
        \centering
        \includegraphics[width=\textwidth]{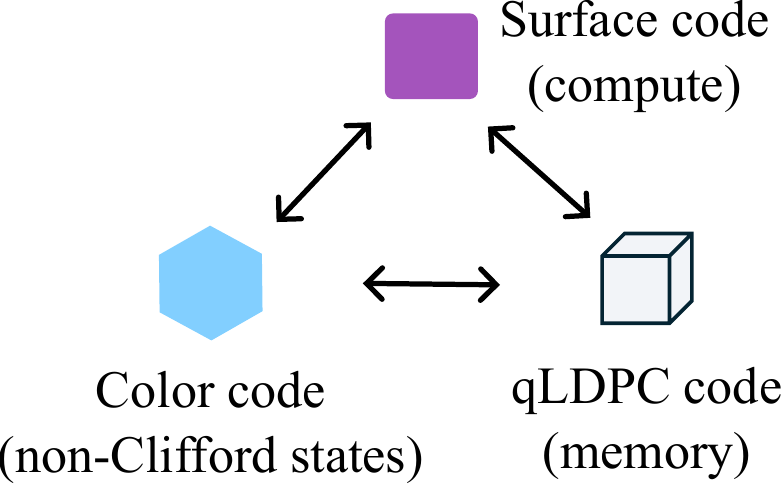}
        \caption{}
        \label{subfig:intro-codes}
    \end{subfigure}
    \hspace*{\fill}
    \begin{subfigure}[b]{0.18\linewidth}
        \centering
        \includegraphics[width=\textwidth]{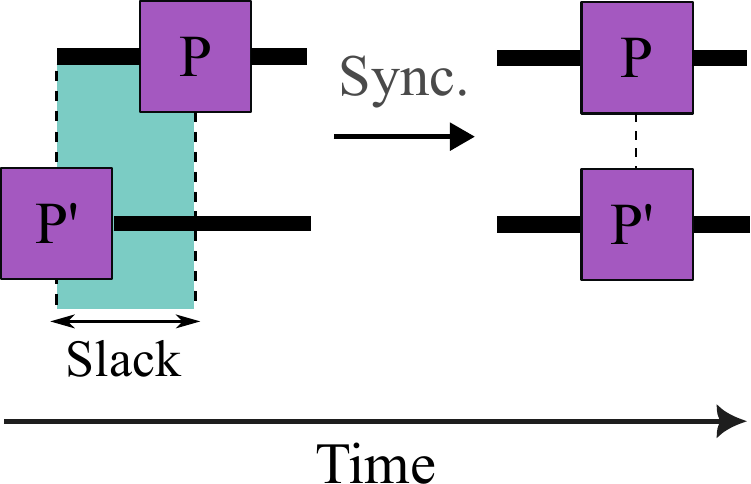}
        \caption{}
        \label{subfig:intro-1}
    \end{subfigure}
    \hspace*{\fill}
    \begin{subfigure}[b]{0.3\linewidth}
        \centering
        \includegraphics[width=\textwidth]{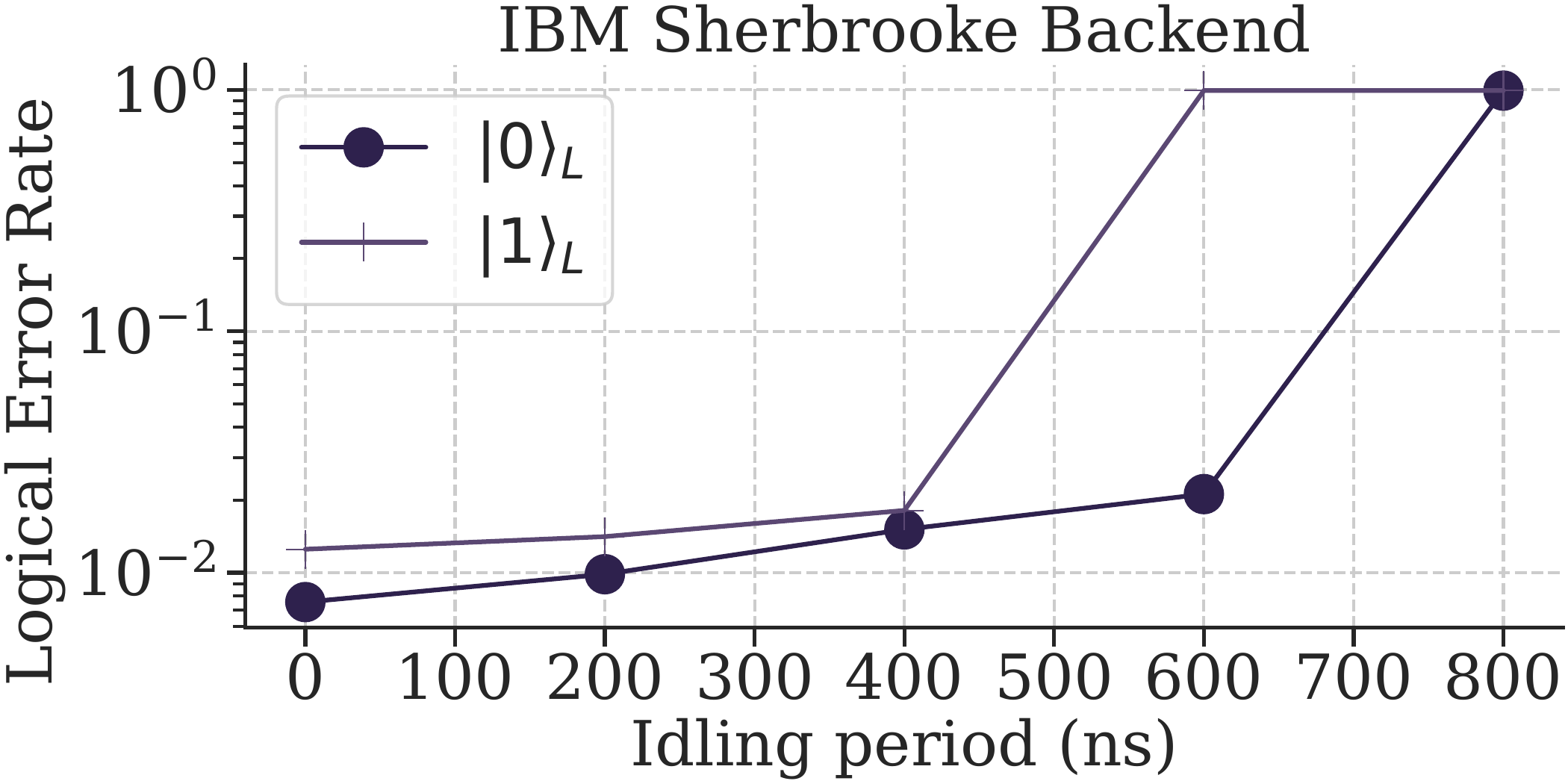}
        \caption{}
        \label{subfig:intro-2}
    \end{subfigure}
    \hspace*{\fill}
    \begin{subfigure}[b]{0.21\linewidth}
        \centering
        \includegraphics[width=\textwidth]{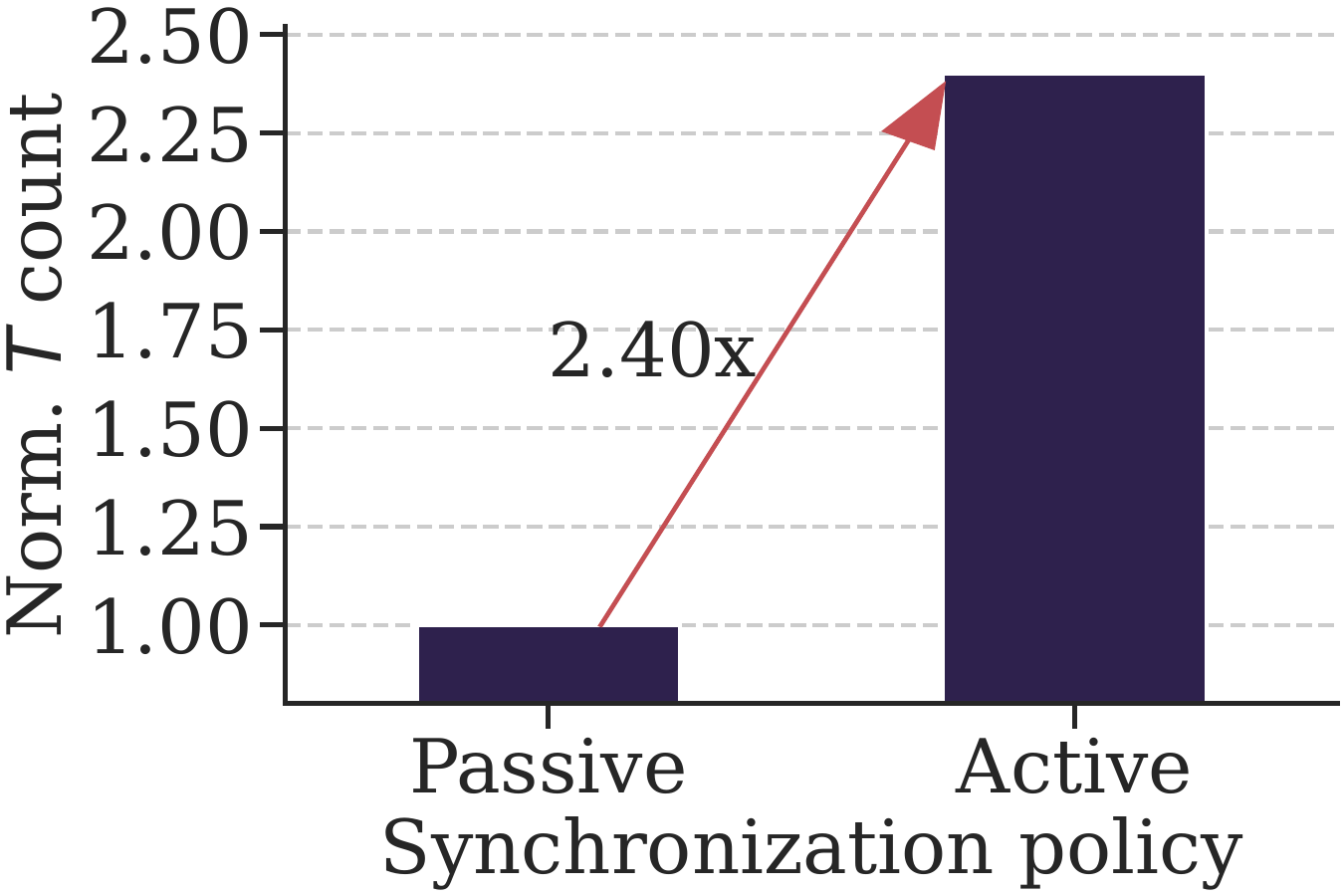}
        \caption{}
        \label{subfig:intro-3}
    \end{subfigure}
    \caption{
    (a) A heterogeneous FTQC system can optimize space-time volume by combining different QEC codes: surface codes excel in computation, qLDPC codes serve well as memory with high coding rates and color codes efficiently produce low-cost non-Clifford (magic) states.
    (b) Patches $P, P'$ are desynchronized if their syndrome generation cycles are in different phases at a given point in time. Synchronizing them ensures both patches start their next syndrome generation cycle at the same time;
    (c) Experimental evaluations on IBM Sherbrooke~\cite{ibm_sherbrooke} for the logical error rate (LER) of the three-qubit repetition code as a function of the idling period added before the second (and final) round of syndrome measurements (both $|0\rangle_L=|000\rangle, |1\rangle_L=|111\rangle$ logical observables). 
    (d) The proposed \textit{Hybrid} and \textit{Active} synchronization policies reduce the LER compared to the Passive policy, enabling deeper circuits.
    }
    \vspace{-0.15in}
    \Description[Introductory figure]{}
    \label{fig:intro}
\end{figure*}
 
The use of Quantum Error Correction (QEC)~\cite{Shor1995} is the most promising step toward practical, Fault-Tolerant Quantum Computing (FTQC). 
By using error-correcting codes such as the surface code~\cite{fowler2012, Knill1998} that encode the state of each logical qubit onto many physical qubits, it is possible to reduce the error rates on quantum computers to run algorithms such as quantum simulation~\cite{Lloyd1996}, factoring~\cite{Shor1997}, and chemistry simulations~\cite{AspuruGuzik2005}. 
Recent experimental demonstrations have already shown the advantage of using such error-correcting codes~\cite{acharya2024quantumerrorcorrectionsurface, GoogleSuppressing, Bluvstein2023, Krinner2022, Ryan2021}. 
In addition to reducing device-level noise, systems must be scaled to support the large number of qubits required by fault-tolerant quantum computers, with the $400+$ qubit system unveiled by IBM~\cite{IBM400} and roadmaps leading to larger systems to support error-correcting codes~\cite{GoogleQuantumRoadmap, IBMQuantumRoadmap} in the near future. \rev{QEC codes work by executing a fixed sequence of gates followed by measurements, which constitute one \textit{syndrome generation cycle}. This process is repeated, and the measurements made are used to detect errors that may have occurred in the system.}
% Despite the significant experimental progress being made to achieve the targets set by these roadmaps, additional system-level studies are necessary for future error-corrected quantum computers. 
In this paper, we focus on the systems-level problem of logical qubit synchronization for the surface code using Lattice Surgery. 
While our evaluations are restricted to patch-based encodings using the surface code and Lattice Surgery due to their well-understood and well-defined properties, the problem we are addressing will hold true for other QEC codes such as color codes~\cite{Fowler2011} and qLDPC codes~\cite{Bravyi2024}.

% \subsection{Logical Operations using Lattice Surgery}
Lattice Surgery has emerged as the most promising candidate for facilitating universal error-corrected quantum computing using the surface code~\cite{horsman2012surface, Litinski2018, fowler2019low, Litinski2019}. In the patch-based encoding used by Lattice Surgery, all logical operations (CNOTs, non-Clifford gates) can be decomposed into \texttt{split} and \texttt{merge} operations. These operations introduce a strict synchronization requirement on all logical patches (qubits) in the system -- in order to \texttt{merge} two (or more) logical patches, the syndrome generation cycle for all patches involved must begin at the same time immediately after the \texttt{merge} operation. If one patch is in the middle of its syndrome generation cycle while the rest are at the beginning of their cycles, the patches cannot be merged immediately.

% \subsection{The Logical Clock and Desynchronization}

What can cause desynchronization among logical patches? Desynchronization occurs when a patch is at a different stage in its syndrome generation cycle compared to others involved in a Lattice Surgery operation. To clarify the role of synchronization in FTQC systems, we introduce the concept of \textbf{logical clocks}, each patch completes its syndrome generation cycle within a single logical clock cycle. However, even with uniform physical gate and measurement latencies, the phase of logical clocks can vary between patches, leading to desynchronization. This variation arises from several sources, outlined below:

% What could cause desynchronization among logical patches? If a patch is in a different phase of its syndrome generation circuit compared to other patches involved in a Lattice Surgery operation, it is effectively desynchronized from the other patches. To understand synchronization and its need in FTQC systems, we first describe the notion of \textbf{logical clocks}. Every patch starts and ends its syndrome generation cycle within a logical clock cycle. Now, even if all operational (gate and measurement) latencies are uniform, the logical clock for different logical patches can be different, resulting in desynchronization. This is because of sources listed below:

\subheading{Heterogeneous FTQC systems}
In a homogeneous system composed of only surface code patches, synchronization will be unnecessary (assuming a uniform, defect-free system). The production of non-Clifford $T$ states (magic states) will be performed in distillation factories composed entirely of surface code patches. However, the surface code suffers from a poor encoding rate, making the physical qubit requirements for practical applications extremely high. To combat this, recent works use qLDPC codes as quantum memories~\cite{Stein2024, Bravyi2024} and use color codes for magic state distillation~\cite{gidney2024cultivation,ito2024, Butt2024, Daguerre2024, PoulsenNautrup2017, Chamberland2020}, which drastically reduce the physical qubit requirements of FTQC systems. These codes can be used in conjunction with the surface code for universal FTQC, as shown in Figure~\ref{subfig:intro-codes}. However, this introduces additional logical clock domains in the system -- color codes and qLDPC codes have a logical clock that is significantly different from the surface code due to differences in the number of CNOTs and flag qubit measurements in the syndrome generation circuits. These differences in the logical clocks will necessitate synchronization between logical qubits, especially when operations span logical clock boundaries.

% {\color{red} - Does qldpc/color code use flag qubits?}

\subheading{Dropouts, defects}
Even in a homogeneous system of surface code patches, synchronization may still be necessary due to fabrication defects or failures in qubits and couplers~\cite{Lin2024, luci2024, Yin2024, McEwen2023, Strikis2023, Auger2017}. Such dropouts and defects can be managed by modifying the syndrome generation circuit, although this adjustment may result in some logical patches experiencing extended clock cycles~\cite{luci2024}, thereby introducing desynchronization across the system.

% Even in a homogeneous system composed of only surface code patches, synchronization might be necessary due to fabrication defects and/or qubit/coupler failures~\cite{Lin2024, luci2024, Yin2024, McEwen2023, Strikis2023, Auger2017}. These dropouts and defects can be tolerated by modifying the syndrome generation circuit, which can lead to some logical patches having longer logical clock cycle times than others, resulting in desynchronization.

\subheading{Other sources}
Speculative execution of leakage reduction circuits~\cite{Vittal2023}, twist-based lattice surgery~\cite{Litinski2019magic, Chamberland2022twist}, and uneven CNOT latencies from chip-to-chip couplers in modular architectures~\cite{Smith2022, McKinney2023, McKinney2024, Rodrigo2020, Sete2021, Field2024} can create variable-length surface code cycles, can also lead to desynchronization in logical patches within a homogeneous system.

Our evaluations show that large-scale fault-tolerant quantum computing (FTQC) systems can not operate in lockstep due to the heterogeneity of codes and imperfect qubit lattice, resulting in non-uniform syndrome generation cycles requiring forced synchronizations. \textit{We now ask: How can we reliably synchronize logical patches?} Figure~\ref{subfig:intro-1} illustrates the need for synchronization between two logical patches, where the leading patch $P$ must either slow down or pause error correction to allow the lagging patch $P'$ to catch up. However, the idling can compromise quantum error correction, causing uncorrectable errors. Recent experimental results from Google reveal that idling errors are one of the primary causes of logical failures ~\cite{acharya2024quantumerrorcorrectionsurface}, with data-qubit idling alone accounting for over 20\% of the total error budget.

To investigate the impact of idling errors on QEC, we implemented a three-qubit repetition code circuit on IBM’s Sherbrooke\footnote{We used qubits 33,37-40 with worst-case $T_1=330.77\mu s$, $T_2=72.68\mu$s.} quantum computer~\cite{ibm_sherbrooke}, introducing idling delays before the final syndrome measurement round. The impact of idling error is evident in Figure~\ref{subfig:intro-2}: logical error rates increase dramatically with idling durations, as idling significantly increases the likelihood of uncorrectable errors due to decoherence, thus sharply reducing logical fidelity. In these trials, we used a look-up table (LUT)-based decoder over 20,000 shots, implementing an 
``X-X" dynamical decoupling sequence during all idle periods to mitigate decoherence.

To reduce the impact of idling errors incurred during synchronization, we introduce an \textit{Active} synchronization policy that proactively distributes the synchronization slack among multiple syndrome generation rounds. \textit{The key insight behind this approach is that smaller delays incur fewer decoherence errors in physical qubits}, and thus result in fewer logical errors. Compared to the baseline \textit{Passive} synchronization approach which idles the leading logical patch for the entire synchronization slack right before the Lattice Surgery operation, the \textit{Active} synchronization policy achieves a reduction of up to 2.4$\times$ in the logical error rate for two surface code patches. This 2.4$\times$ reduction in the logical error rate can in turn allow deeper logical circuits with more non-Clifford $T$ gates, as shown in Figure~\ref{subfig:intro-3}. In cases where the syndrome generation cycle times of the patches are unequal, synchronization can also be achieved by running error correction for additional rounds. We call this the \textit{Extra Rounds} policy, and we augment it  with \textit{Active} synchronization to create the \textit{Hybrid policy} that \textbf{reduces the LER by up to 3.4$\times$}. 
Furthermore, we also show that by reducing the logical error rate as compared to the \textit{Passive} policy, the decoding latency per Lattice Surgery operation can be reduced by up to 2.2$\times$.

Our work is the first to focus on synchronization policies for FTQC systems. We have developed a scalable and versatile simulator~\cite{latticesim} to study the impact of non-uniform gate latencies and synchronization overheads on the logical error rate and performance. 
The primary contributions of this work are as follows:

\begin{itemize}[topsep=0pt, leftmargin=*]
    \item We show the necessity of synchronization and how idling during synchronization can substantially increase logical error rates.

\item We propose \textit{Active} synchronization and show how proactive synchronization improves the logical error rate.

\item We propose the \textit{Hybrid} policy that can further improve the performance of the \textit{Active} policy.

% \item We motivate the \textit{Active} policy with real experimental data from IBM systems.

\item We release an open-source stabilizer circuit generator integrated with Stim that can simulate surface code Lattice Surgery. 
\end{itemize}

\section{Background}
\label{sec:background}

In this section, we discuss relevant background related to the surface code and Lattice Surgery.

% \subsection{A High-Level View of FTQC}

% We envision a layered architecture for quantum computers running surface codes to enable fault-tolerance~\cite{cody}. At the top, we have a quantum algorithm, e.g., Shor's algorithm, which can be compiled into a sequence of logical operations applied to logical data qubits. A logical qubit is a collection of physical qubits, and by continuously running QEC, they are made significantly more reliable than individual physical qubits. 
% % Logical operations can be further decomposed into operations such as merge, split, and measurement operations. 
% At the lowest level, we continuously run QEC cycles that entangle data and measure qubits to compute parity to detect errors. 
% \emph{QEC cycles across all logical qubits must be synchronized to ensure correct multi-qubit logical operations}.

\subsection{Surface Code and Lattice Surgery}

\begin{figure}[t]
    \centering
    \hspace*{\fill}
    \begin{subfigure}[b]{0.4\linewidth}
        \centering
        \includegraphics[width=\textwidth]{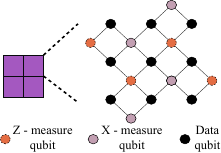}
        \caption{}
        \label{subfig:bg_patch}
    \end{subfigure}
    \hspace*{\fill}
    \begin{subfigure}[b]{0.22\linewidth}
        \centering
        \includegraphics[width=\textwidth]{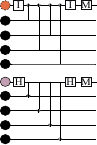}
        \caption{}
        \label{subfig:bg_ckt}
    \end{subfigure}
    \hspace*{\fill}
    \begin{subfigure}[b]{0.31\linewidth}
        \centering
        \includegraphics[width=\textwidth]{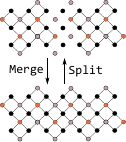}
        \caption{}
        \label{subfig:bg_ls}
    \end{subfigure}
    \hspace*{\fill}
    \\
    \hspace*{\fill}
    \begin{subfigure}[b]{0.45\linewidth}
        \centering
        \includegraphics[width=\textwidth]{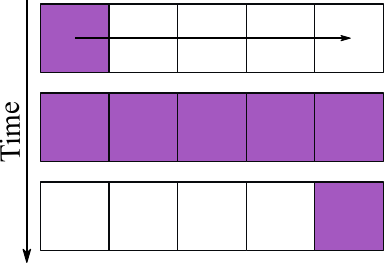}
        \caption{}
        \label{subfig:ls1}
    \end{subfigure}
    \hspace*{\fill}
    \begin{subfigure}[b]{0.3\linewidth}
        \centering
        \includegraphics[width=\textwidth]{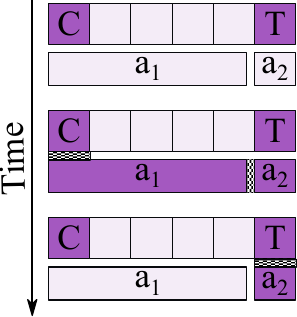}
        \caption{}
        \label{subfig:ls2}
    \end{subfigure}
    \hspace*{\fill}
    \caption{
    (a) A patch in the lattice consists of a grid of physical qubits that are either measure or data qubits; 
    (b) \rev{A syndrome generation cycle: A sequence of gates followed by a measurement is performed on all physical qubits in a patch}; 
    (c) Two patches can be split or merged; 
    (d) An example of patch movement -- the intermediate step involves merging the empty patches between the source and destination patches; 
    (e) Long-range CNOTs use ancillary qubits that span the distance between the control and target logical qubits.}
    \vspace{-0.1in}
    \Description[Background figure]{}
    \label{fig:surface}
    
\end{figure}

%% Cite all experimental demos of surface code  
The surface code is a leading candidate for implementing FTQC systems~\cite{bravyi1998quantum, fowler2012}. Figure~\ref{subfig:bg_patch} shows how a logical patch (qubit) is composed of a fixed arrangement of physical qubits. The physical qubits are categorized as data qubits, X-measure, and Z-measure qubits. The number of X-, Z-measure, and data qubits needed to define a single logical qubit is dependent on the code distance of the surface code, which is a measure of the number of errors that can be corrected. In this work, we assume the use of the rotated lattice~\cite{horsman2012surface}, which consists of $d^2$ data qubits and $d^2 - 1$ measure qubits for a code distance of $d$.  A logical qubit with code distance $d$ can correct $\frac{d-1}{2}$  errors by repeating $d$ rounds of the stabilizer circuit. Figure~\ref{subfig:bg_ckt} shows the circuits executed to generate syndromes for a single round of error correction for both $X$ and $Z$ parity qubits.

% \subheading{Lockstep Execution of Stabilizer Circuit}
% Figure~\ref{fig:surface}(b) shows the interactions between the data qubits and measure qubits for the surface code. For every X- and Z-measure qubit in the patch, all gate operations must be performed in lockstep~\cite{fowler2012}, such that  gates across a large physical lattice are executed in parallel for all measure qubits in a patch -- this level of concurrency must be supported to prevent qubits from idling and thus incurring errors. A surface code cycle ends with the measurement of the measure qubits, and multiple code cycles are needed to detect errors accurately. 

% It is essential to minimize the idling qubits.  

\subheading{Lattice Surgery}
Figure~\ref{subfig:bg_ls} shows two fundamental Lattice Surgery operations -- the \texttt{merge} and \texttt{split} operations. The \texttt{merge} operation fuses two logical qubits to produce a single logical qubit whose state depends on the states of the two logical qubits that were merged. 
% Note that a merge introduces new interactions between the measure and data qubits of the two patches, and these interactions have to be synchronized with the other interactions in the two patches. 
\rev{During a \texttt{merge}, the physical qubits in the buffer between the two patches become a part of both patches, and the gate schedule of Figure~\ref{subfig:bg_ckt} across both patches must then start in lockstep, as it does within a single patch.} The \texttt{split} operation splits a logical qubit into two. \emph{All logical multi-qubit operations can be decomposed into a sequence of merge, split, and measurement operations.}

%% 
% \subsection{Hardware Architectures and Experimental QEC}
% \label{sec:background_qec}

% \begin{figure}[t]
%     \centering
%     \includegraphics[width=\linewidth]{figs/background.pdf}
%     \caption{A control architecture for a large lattice of qubits will require an array of controllers. Within a controller, the control and readout pipelines are used to execute the syndrome circuits and measure the qubits respectively.}
%     \label{fig:ctrl_hardware}
% \end{figure}
% \begin{figure}[b]
%     \centering
    
% \includegraphics[width=0.9\linewidth]{figs/perror.pdf}
    
%     \caption{ Operational error rates for Google hardware. Original data reported in Table-I  of supplementary material~\cite{GoogleSuppressing}.}
%     \Description[bg figure]{}
%     \label{fig:google_data}
% \end{figure}

\begin{figure*}
    \centering
    % \hspace*{\fill}
    \begin{subfigure}[b]{0.27\linewidth}
        \centering
        \includegraphics[width=\textwidth]{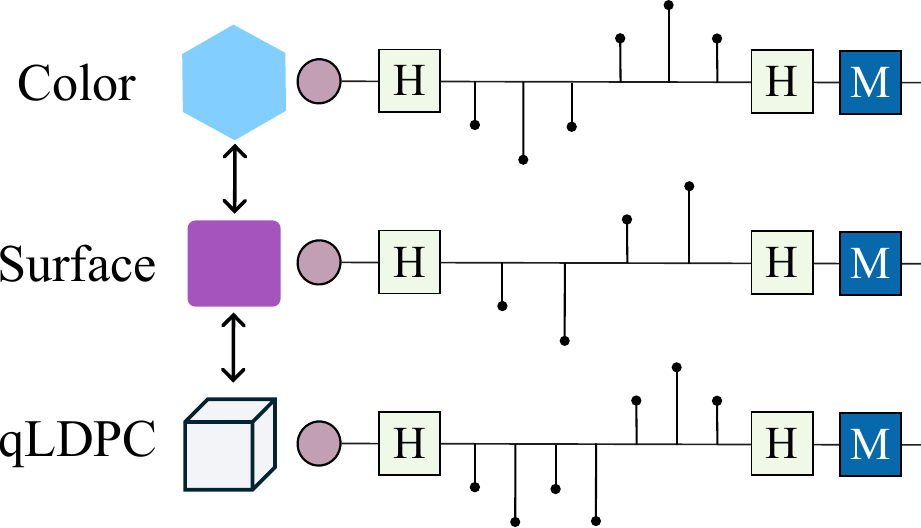}
        \caption{}
        \label{subfig:motive-1}
    \end{subfigure}
    % \hspace*{\fill}
    \hfill
    \begin{subfigure}[b]{0.3\linewidth}
        \centering
        \includegraphics[width=\textwidth]{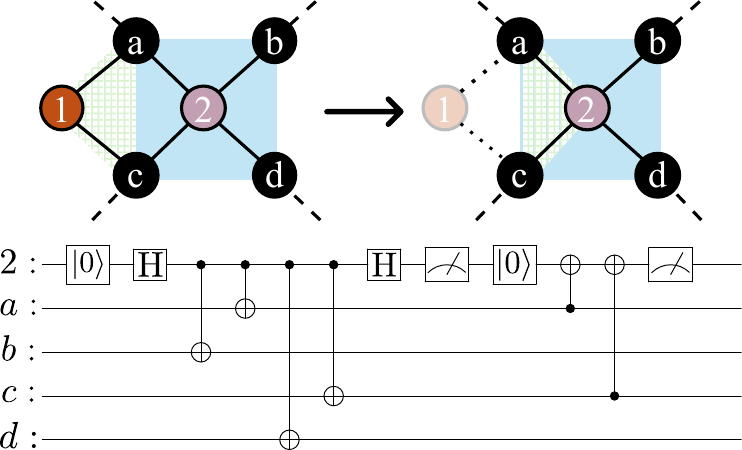}
        \caption{}
        \label{subfig:motive-2}
    \end{subfigure}
    \hfill
    % \hspace*{\fill}
    % \begin{subfigure}[b]{0.33\linewidth}
    %     \centering
    %     \includegraphics[width=\textwidth]{figs/problem-3.pdf}
    %     \caption{}
    %     \label{subfig:motive-3}
    % \end{subfigure}
    \begin{subfigure}[b]{0.34\linewidth}
        \centering
        \includegraphics[width=\textwidth]{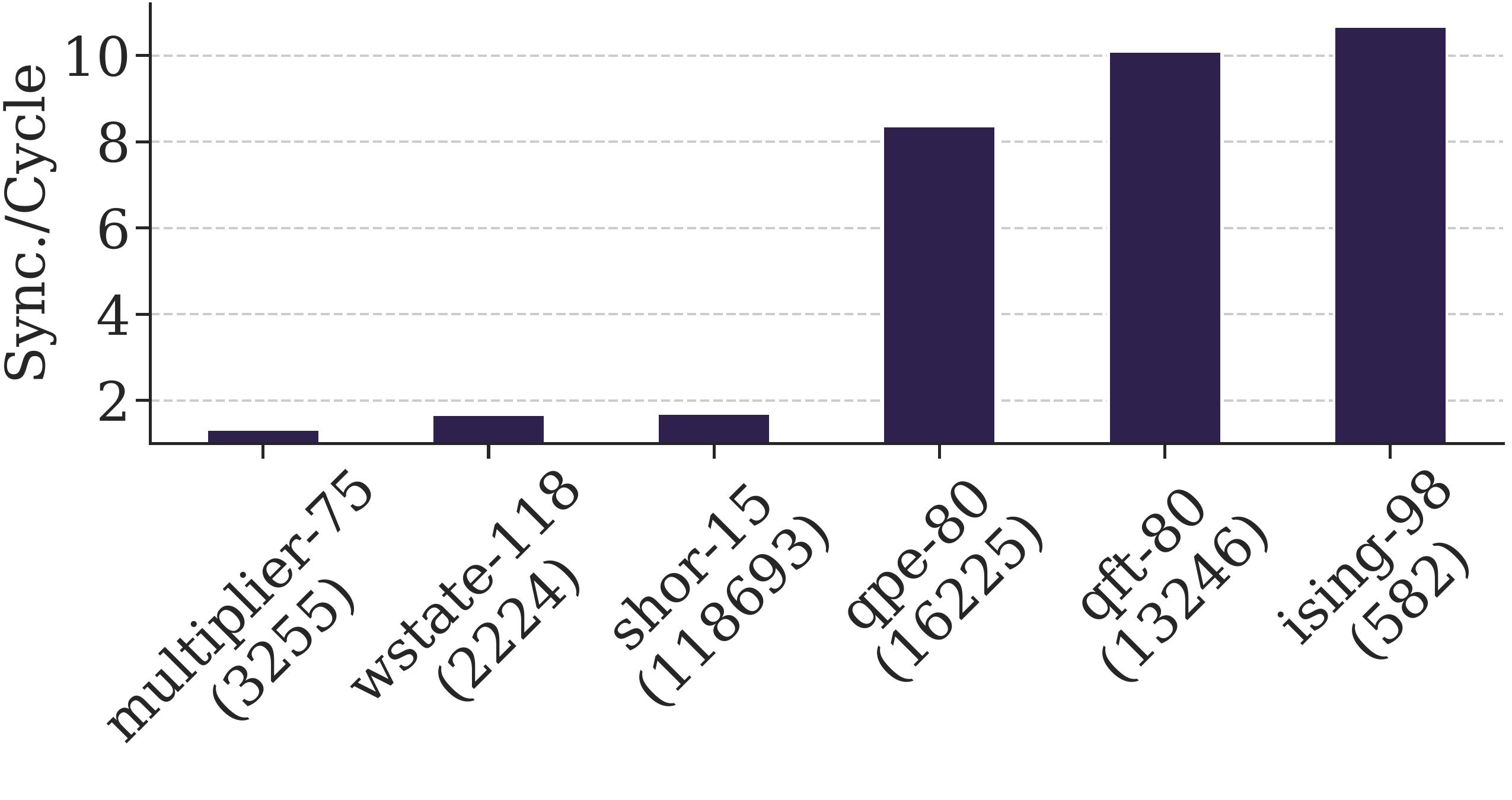}
        \caption{}
        \label{subfig:motive-4}
    \end{subfigure}
    \vspace{-0.05in}
    \caption{
    (a) Teleporting logical qubits between different codes will require synchronization since every code has a different number of CNOTs in its syndrome generation cycle and hence has a different cycle duration;
    (b) Dropouts result in some qubits and/or coupler links being unusable (top). This can be circumvented by time-multiplexing of some qubits to construct syndrome generation circuits that have a duration that is longer than (but not a multiple of) the original duration (bottom);
    (c) Estimate of the minimum number of synchronizations per logical cycle (total logical cycles required annotated).
    % {\color{red}{change (c) to show frequency of T gates = Tgate/total gates, change (d) number of synchronizations/total surface code rounds }}
    % {\color{red} make (a) smaller and (b) larger in size, highlight the merge joint, (a) has less info, so less real estate}
    }
    \vspace{-0.1in}
    \Description[Motivational figure]{}
    \label{fig:motive}
    
\end{figure*}

\subsection{Logical Operations Requiring Synchronization}

 % Quantum computers running surface code are envisioned to have a ``sea of qubits architecture", where a lattice of physical qubits is partitioned into patches of logical qubits. 
 There are two types of qubit patches: logical data patches and logical ancillary patches. Data qubit patches hold data variables described in the algorithm, while ancillary patches facilitate logical operations. Figure~\ref{subfig:ls1} shows a total of five patches. The filled patch in the top panel corresponds to data qubits, and unfilled (in white color) patches are ancillary qubits. All FTQC algorithms involve operations that span multiple patches. For performing multi-patch operations, patches must be synchronized. There are three broad categories of multi-patch operations.
 
\subsubsection{Moving patches} Lattice Surgery allows the movement of logical qubits across the lattice. As shown in Figure~\ref{subfig:ls1}, the movement of a patch involves merging the logical qubit to expand it with the empty ancillary patches till the destination patch. This expanded patch is measured for $d$ code cycles, after which the extra qubits are measured out. 
% Patches might need to be moved to enable other logical operations like CNOTs, for moving magic states to a region where they can be consumed by other logical qubits or to move a patch away from a more error-prone region of the lattice. 
% Since the expanded patch is composed of smaller patches that may or may not have the same controller, it is necessary for the patches to be synchronized during the operation. 

\subsubsection{Long-range CNOTs} CNOTs are essential for implementing any quantum program and are also used frequently for consuming $T$ states. Often, logical patches involved are not in the same vicinity, necessitating long-range CNOTs that span multiple patches in the lattice. Figure~\ref{subfig:ls2} shows how a long-range CNOT can be performed between the Control (C) patch and the Target (T) patch, which are separated by numerous patches by using Lattice Surgery. The two ancillary qubits are first merged with the Control patch, after which the second ancillary is merged with the Target patch~\cite{Litinski2019}.

\subsubsection{Non-Clifford gates} 
Non-Clifford gates are necessary for building universal FTQC systems. The application of non-Clifford gates also involves multi-patch operations in the form of Pauli-product measurements~\cite{Litinski2018, Litinski2019, fowler2019low} using Lattice Surgery. Thus, every non-Clifford operation in a program will require a synchronized Lattice Surgery operation between at least two patches.
% Magic State Distillation is the most resource and time intensive operation in a fault-tolerant quantum computer.
% Magic State Distillation is used to prepare high-quality \emph{magic} states, purified forms of the non-Clifford $T$ state. Distillation is performed in dedicated regions of the lattice called the $T$ factories, which continuously produce magic states for consumption by the algorithmic logical qubits. 
% Not only do distillation protocols utilize patch movement and long-range CNOTs described above to produce magic states, the consumption of magic states also involves multi-patch operations in the form of Pauli-product measurements~\cite{Litinski2018, Litinski2019, fowler2019low}.

\subsection{Idling and its Impact}
Dynamical Decoupling (DD)~\cite{Khodjasteh2005, Khodjasteh2007, Bylander2011, Das2021} has been shown to be an effective mitigation against qubit decoherence due to idling. Despite the use of DD in recent QEC experiments by Google, qubit idling is still the second-highest contributor to their overall error budget~\cite{acharya2024quantumerrorcorrectionsurface, GoogleSuppressing}. Limited effectiveness of DD is observed on LBNL's AQT platform as well, the idling error with DD is around 0.1\% per 50 ns window~\cite{rudinger2021experimental}. \textbf{DD is thus merely a mitigation against idling errors, and in this work, we aim to augment its effectiveness.}

\section{Motivation for Synchronization}
\label{sec:motivation}

% TODO:
% Motivate intra patch scheduling (runahead/fenced)

%This section proposes a \emph{Locally Synchronous Globally Asynchronous (LSGA)} Execution Model. First, we motivate the need for synchronization primitives and then discuss the high-level model.
In this section, we highlight the need for synchronization primitives and how often they will be needed.

\subsection{When is Synchronization not needed?}
For a system composed of perfect couplers and qubits with a homogenous QEC code, synchronization will not be necessary since all operations will occur in lockstep. However, due to the poor coding rate of the surface code, using a heterogeneous mix of QEC codes will yield better space-time trade-offs in future systems~\cite{Stein2024}. However, this heterogeneity, along with other factors, will result in logical qubits being desynchronized from one another. 

\subsection{Sources of Desynchronization}
% We now discuss the different sources of desynchronization that could exist in a FTQC system. 

% \todo{add figures and figure references}

\subsubsection{Using better codes for non-Clifford gates}
Non-Clifford gates are a crucial component for universal FTQC. Distillation of non-Clifford $T$ states using triorthogonal codes \cite{Litinski2019magic} was up until recently the most promising means of producing high-quality magic states for the surface code. However, recent work introducing magic state cultivation~\cite{gidney2024cultivation} has been shown to be capable of producing good quality magic states for a fraction of the space-time cost. However, these constructions, along with many others~\cite{Chamberland2020, ito2024, Butt2024, Daguerre2024, PoulsenNautrup2017}, use different codes and rely on teleporting logical qubits between a code that is better for producing magic states and the surface code. This automatically introduces desynchronization between patches involved in magic state production and other algorithmic logical qubits in the system, since the syndrome generation cycle time for color/other codes is different from the surface code, as shown in Figure~\ref{subfig:motive-1}. Quantum LDPC codes~\cite{Bravyi2024, Stein2024} also rely on teleporting logical qubits to the surface code for executing non-Clifford gates, resulting in desynchronization.
% \footnote{Desynchronization causes logical clocks to fall out of phase.}

\subsubsection{Circuit construction in the presence of dropouts}
Exploiting time-dynamics to construct error correction circuits that can tolerate defects/failed couplers and qubits in a lattice (called dropouts) have been proposed~\cite{luci2024, McEwen2023, Yin2024}. These works are part of a broader set of works that seek to achieve fault-tolerance even in the presence of fabrication defects~\cite{Strikis2023, Lin2024, Siegel2023, Auger2017}. As shown in Figure~\ref{subfig:motive-2}, these constructions can result in patches with syndrome generation cycle times that are longer than other patches in the system, thus introducing desynchronization.

\subsubsection{Twist-based Lattice Surgery}
Measurements involving Pauli $Y$ operators benefit greatly from twist-based Lattice Surgery~\cite{Litinski2018, Chamberland2022twist}. However, patches enabling twist-based Lattice Surgery on the surface code require additional CNOTs in the {syndrome generation} circuit, which will desynchronize them with regular surface code patches in the system. 

\subsubsection{Modular architectures}
Industrial roadmaps~\cite{IBMQuantumRoadmap, GoogleQuantumRoadmap} and several works have explored chiplet-based modular architectures~\cite{Smith2022, McKinney2023, McKinney2024, Rodrigo2020, Zhang2024} and chip-to-chip couplers~\cite{Sete2021, Field2024} for implementing such modular architectures. However, movement of patches or logical operations across chiplet boundaries could result in desynchronization of those patches from the remaining system due to higher CNOT latencies at the boundary of chiplets. 

% \subsubsection{Other sources}
% % So far, the desynchronization sources discussed exist despite all gate and measurement latencies being uniform. However, if this is not the case, logical patches can get desynchronized. 
% Speculative execution of leakage reduction circuits (LRCs), which insert additional physical operations non-deterministically~\cite{Vittal2023}, can lead to varying cycle durations across patches, causing desynchronization.

\subsection{Lower Bound on Synchronizations Required}
How often will synchronized Lattice Surgery operations be required on a FTQC system? To understand a rough lower bound on the number of synchronized operations needed for different workloads, we used the Azure QRE toolbox~\cite{beverland2022assessing} for determining the number of magic states required for those workloads. Since magic state consumption requires at least one Lattice Surgery operation~\cite{Litinski2019}, we can determine the frequency of such synchronized Lattice Surgery operations by dividing the number of magic states needed by the total number of error correction cycles needed to execute a program. This can give us a rough lower bound on the number of synchronizations required, especially for systems using different QEC codes for computation and $T$ state production. As shown in Figure~\ref{subfig:motive-4}, {about 1 to 11 operations requiring synchronization will be performed every error correction cycle}, thus highlighting the need for having reliable policies for synchronizing logical qubits -- the frequency is highly dependent on the parallelism exhibited by an application.

\begin{figure}[b]
    \centering
    \vspace{-0.2in}
    % \hspace*{\fill}
    \begin{subfigure}[b]{0.49\linewidth}
        \centering
        \includegraphics[width=\textwidth]{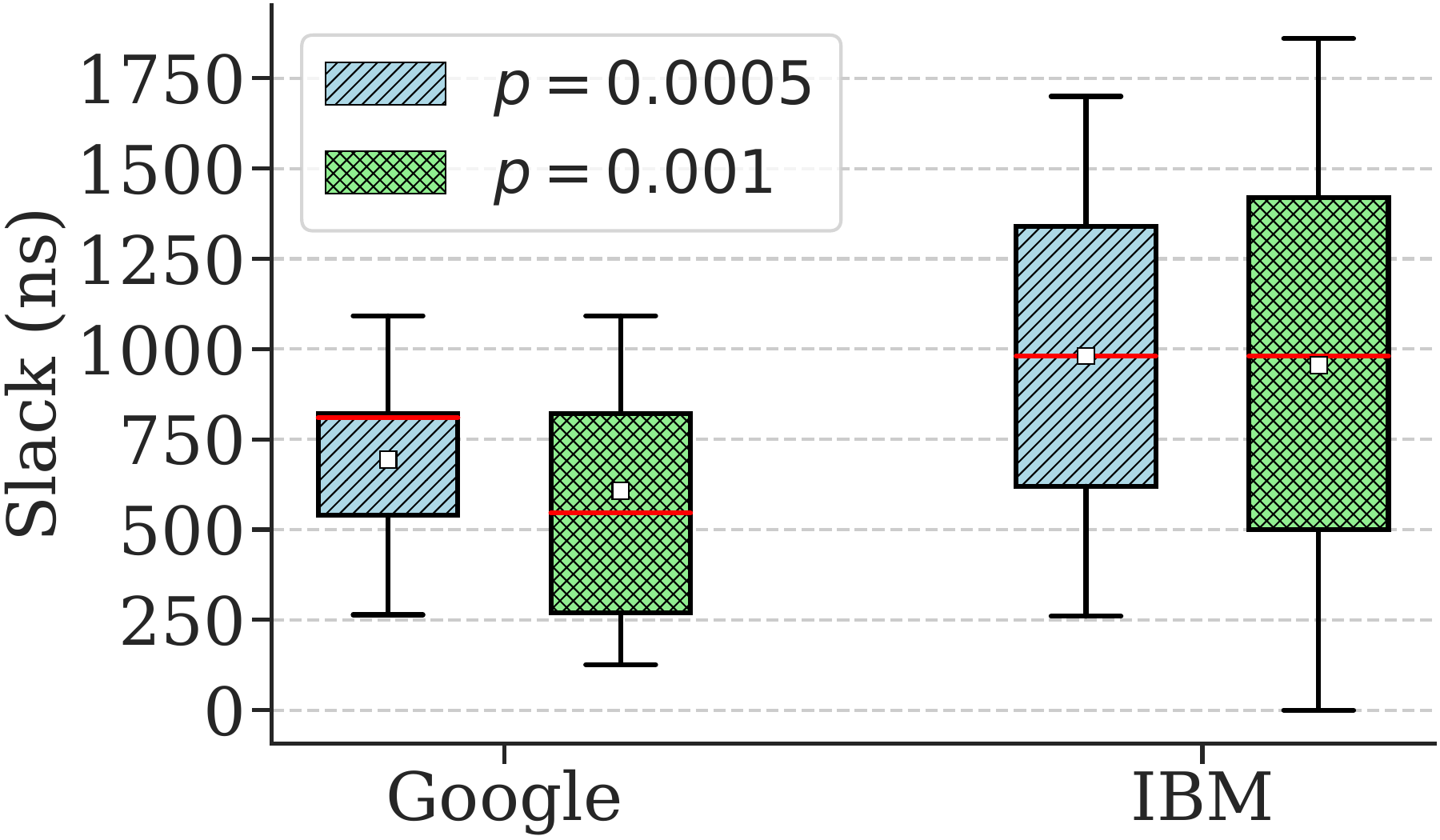}
        \caption{}
        \label{fig:cult_slack}
    \end{subfigure}
    \hspace*{\fill}
    \begin{subfigure}[b]{0.49\linewidth}
        \centering
        \includegraphics[width=\textwidth]{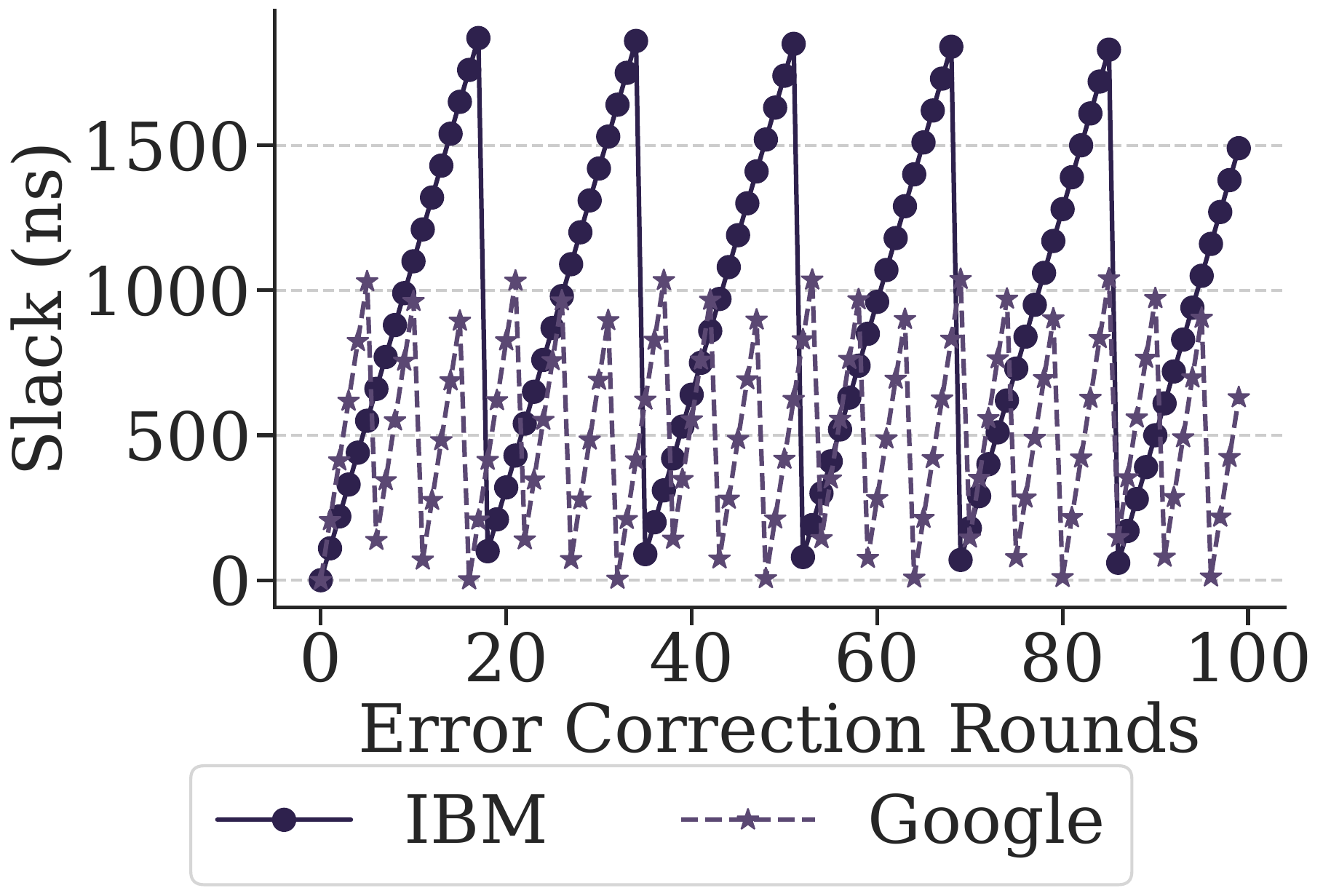}
        \caption{}
        \label{fig:qldpc_slack}
    \end{subfigure}
    % \hspace*{\fill}
    \caption{
    \rev{
    (a) Slack distribution for IBM and Google-like systems when magic state cultivation~\cite{gidney2024cultivation} is used. The square represents the mean;
    (b) Slack as a function of error correction rounds when qLDPC codes are used as memories with surface code patches for IBM and Google systems (independent of $p$). 
    }
    }
    \Description[Hamm wt.]{}
    \label{fig:slack_motivation}
\end{figure}

% \vspace{-0.2in}
\rev{
\subsection{Case Studies}
\label{sec:motive_case_studies}
We now discuss two concrete examples that introduce desynchronization in an FTQC system.
}

\rev{\subsubsection{Magic state cultivation}
Magic state cultivation efficiently prepares high-fidelity T states by incrementally increasing their fault distance within a surface code, outperforming traditional distillation~\cite{gidney2024cultivation}. However, it is inherently non-deterministic, and the final T state may become desynchronized with the surface code patches used for computation. This phase misalignment, or slack, is influenced by the success rate of the cultivation protocol, which in turn depends on the physical error rate $p$, as the number of retries is primarily dictated by $p
$~\cite{gidney2024cultivation}.}

\rev{
To quantify the desynchronization caused by cultivation, we use data for different physical error rates to determine the average and worst-case slack between a patch that yields a non-Clifford state via cultivation and a surface code patch that will use this non-Clifford state. We assume both patches are synchronized in the beginning and simulate this system for 100,000 shots for different physical error rates ($p$). Figure~\ref{fig:cult_slack} shows the median (line) and mean (small square) slacks introduced due to cultivation in a system with IBM and Google gate and measurement latencies (Table~\ref{tab:configs} shows the values used). From Figure~\ref{fig:cult_slack}, we assume the slack to be 500/1000ns for all evaluations since the average case and worst-case correspond to these values.}

\rev{
\subsubsection{Using qLDPC codes}
The use of qLDPC codes for memory and the surface code for computation is a promising hybrid approach~\cite{Stein2024}. To switch between qLDPC and surface code, we can teleport data between two logical qubits of the two codes. This requires perfectly aligned QEC cycles (synchronized) between the two logical qubits. Unfortunately, these codes have different QEC cycle lengths—while the surface code requires 4 CNOT layers per cycle, qLDPC codes need 7 CNOT layers~\cite{Bravyi2024}. This difference causes desynchronization. Furthermore, the slack dynamically changes when teleporting logical qubits between the two codes, as the number of elapsed cycles determines the phase mismatch. Figure~\ref{fig:qldpc_slack} shows this slack as a function of rounds for a surface code and qLDPC logical qubit with Google and IBM architectures.}

% Using qLDPC codes as memories and the surface code for computation has been explored in prior works~\cite{Stein2024}. Due to the difference in syndrome generation cycle times~\cite{Bravyi2024}, teleporting information between two logical qubits of the two codes will require synchronization, since the slack is dependent on the number of cycles that have elapsed since the last synchronization. Figure~\ref{fig:qldpc_slack} shows this slack as a function of rounds for a surface code and qLDPC logical qubit (7 CNOT layers~\cite{Bravyi2024}) that were initially synchronized.

\rev{
The case studies discussed above show that synchronization will be necessary in future FTQC systems, especially ones using heterogeneous architectures. In subsequent sections, we introduce synchronization policies that can enable systems reliably.
}
% closing statements II
% We approach this question at two different abstraction levels -- the first is the synchronization of the patches for enabling multi-patch Lattice Surgery operations. The second abstraction level is within a patch where we address the scheduling of instructions that form the stabilizer circuit. 
% In Section~\ref{sec:synchronization}, we present \textit{Active} synchronization that improves the reliability of synchronizing logical patches involved in multi-patch operations. 
% In Section~\ref{sec:scheduling}, we present \texttt{runahead} scheduling which helps in reducing unnecessary idle periods (and hence decoherence) in the instruction schedule within a patch. 

\section{Synchronization for Logical Operations}
\label{sec:synchronization}
% {\color{red} - Todo for Swamit - Make subsection titles punchy}
% Multi-patch operation require all patches involved to be in sync -- every patch involved must start the merged syndrome cycle at about the same time
% Reasons this is non-trivial -- controller skew, difference in gate latencies between different patches, and decoding latency is non-deterministic. The last factor will exist even if everything else is perfect, thus necessitating a synchronization mechanism.
% Synchronization policies -- make patches wait for slowest to finish / introduce small delays within the surface code cycle for the faster patches 
%       - Different policies for runahead and fenced scheduling within the patches.

% Metric: Number of rounds required for re-synchronizing patches involved in a multi-patch operation
% Number of rounds important, can't be too high because the surface code cannot survive for an indefinite number of rounds
% Simple resynchronization policy vs divide and conquer? 

In this section, we describe how patches can be synchronized to facilitate multi-patch operations.

\begin{figure}[t]
    \centering
    % \hspace*{\fill}
    \begin{subfigure}[b]{0.75\linewidth}
        \centering
        \includegraphics[width=\textwidth]{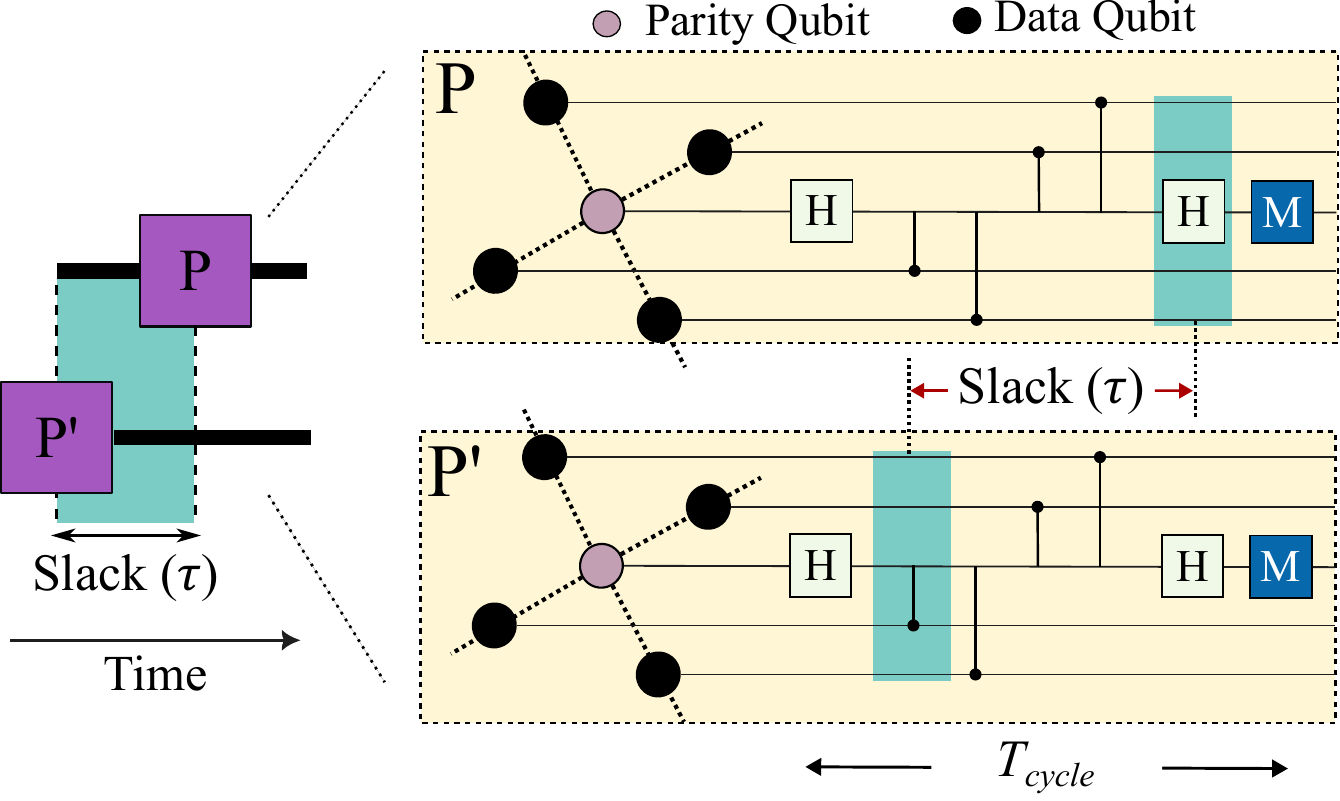}
        \caption{}
        \label{subfig:design_slack}
    \end{subfigure}
    \\
    % \hspace*{\fill}
    \begin{subfigure}[b]{\linewidth}
        \centering
        \includegraphics[width=\textwidth]{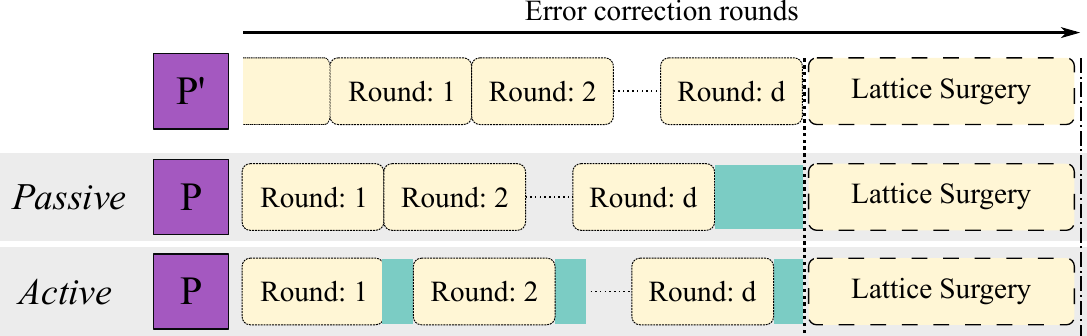}
        \caption{}
        \label{fig:Active_Passive_sync}
    \end{subfigure}

    \caption{
    (a) Patches $P$ and $P'$ are desynchronized by a slack of $\tau$. This corresponds to both patches being in different phases of their syndrome generation cycles (of duration $T_{cycle} = T_P, T_P'$); 
    (b) $P$ and $P'$ can be synchronized by idling $P$ for a period equal to $\tau$ right before Lattice Surgery (\textit{Passive}), or by breaking $\tau$ into smaller segments and adding those idle periods after every round (\textit{Active}) before Lattice Surgery.
    }
    \vspace{-0.1in}
    \Description[Design]{}
    \label{fig:design_act_pass}
\end{figure}

\subsection{Synchronizing Two Patches}
\label{sec:sync_two}

Two patches are desynchronized if they are at different points of their syndrome generation cycle, as shown in Figure~\ref{subfig:design_slack}. Patch $P$ is ahead in time as it is executing the final Hadamard gate, while patch $P'$ is still on its first CNOT, leading to a synchronization slack $\tau$. Note that since $\tau$ is the phase difference in the syndrome generation cycles of $P, P'$, it is bounded by cycle time $T_{cycle}$ since $\tau = \tau~\%~T_{cycle}$. This slack must be eliminated before Lattice Surgery. We now discuss policies that can synchronize patches $P$, $P'$. 

\subsubsection{\textit{Passive} synchronization}
The simplest way of synchronizing two patches would be to have the faster (leading) patch $P$ wait for the slower (lagging) patch $P'$ before starting the Lattice Surgery operation, as shown in Figure~\ref{fig:Active_Passive_sync}. This would essentially pause syndrome generation for patch $P$, thus exposing all data qubits in $P$ to more decoherence errors. While techniques like Dynamical Decoupling (DD)~\cite{Khodjasteh2005, Khodjasteh2007, Bylander2011, Das2021} can be used to reduce the impact of idling (decoherence) errors, they can only partially mitigate such errors, not eradicate them. Secondly, DD sequences are physically applied on hardware and are not error-free -- this reduces their efficacy in reducing idling errors. Recent work by Google~\cite{GoogleSuppressing} shows how idling errors constitute a significant portion of the total error budget \textbf{despite} the use of DD. Finally, DD sequences are calibrated per data qubit~\cite{GoogleSuppressing} to optimize performance, which is not scalable as we increase the number of logical qubits.

\subsubsection{\textit{Active} synchronization}

As an alternative to this \textit{Passive} synchronization approach, we propose \textit{Active} synchronization that achieves synchronization by \textbf{gradually} slowing down the leading logical qubit. Since logical operations are generally performed at the cadence of $d$ rounds~\cite{fowler2012, Litinski2019, fowler2019low}, the leading logical qubit can be slowed down by inserting idle periods before the start of (or after the end of) every round. Doing so breaks the total synchronization slack into smaller fragments and results in fewer errors occurring due to qubit idling, as shown in Figure~\ref{fig:Active_Passive_sync}.
% \rev{Our approach can be co-designed with DD sequences since breaking the synchronization slack into shorter durations that exactly match the DD duration to reduce the impact of idling errors as longer idle periods accumulate more errors even with the use of DD~\cite{Ezzell2023, Das2021}.} 

\begin{figure}[t]
    \centering
    \begin{subfigure}[b]{0.9\linewidth}
        \centering
        \includegraphics[width=\textwidth]{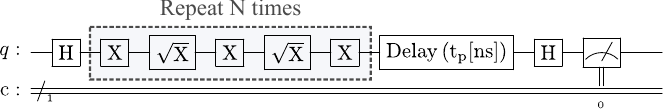}
        \caption{}
        \label{subfig:ibm_Passive}
    \end{subfigure}
    \\
    \begin{subfigure}[b]{0.9\linewidth}
        \centering
        \includegraphics[width=\textwidth]{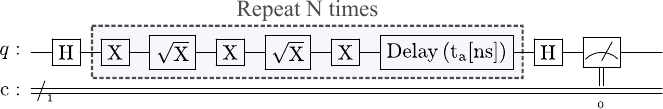}
        \caption{}
        \label{subfig:ibm_Active}
    \end{subfigure}
    \\
    \begin{subfigure}[b]{\linewidth}
        \centering
        \includegraphics[width=\textwidth]{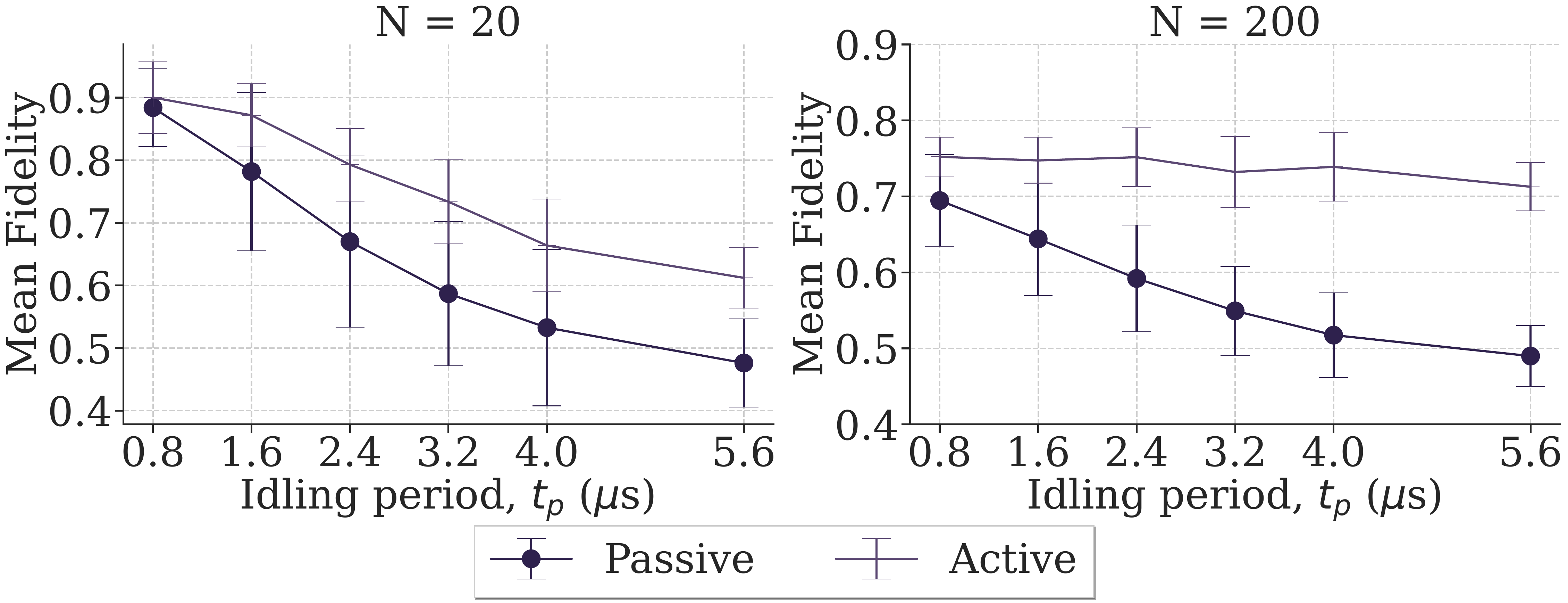}
        \caption{}
        \label{subfig:ibm_charac_ap}
    \end{subfigure}
    \caption{
    Circuits used to emulate the (a) \textit{Passive} and (b) \textit{Active} policies. The delay $t_a$ is repeated $N$ times to yield a net delay of $t_p$, resulting in both circuits having the same total runtime. Both circuits were run on 20 qubits on IBM Brisbane~\cite{ibm_brisbane} for 20,000 shots.
    (c) Experimental mean fidelity across all 20 qubits for $N=20,200$ with different values of $t_p (t_a=t_p/N)$. 
    }
    \vspace{-0.1in}
    \Description[IBM exp figure]{}
    \label{fig:ibm_Active_Passive_exp}
\end{figure}

% \begin{figure}[t]
%     \centering
%     \includegraphics[width=\linewidth]{figs/Active_Passive.pdf}
    
%     \caption{Example showing the difference between \textit{Active} and \textit{Passive} synchronization -- patch $P$ is run for $d$ rounds while the slack is added between (\textit{Active}) or after all (\textit{Passive}) rounds. $P$ is then run for another $d$ rounds before decoding.}
%     \Description[a figure]{}
%     \label{fig:Active_Passive_sync}
    
% \end{figure}

\begin{center}
    \textbf{\emph{\textit{Active} synchronization attempts to decrease the impact of idling errors on the logical error rate by inserting the synchronization slack between multiple code cycles.}}
\end{center}

% Algorithm~\ref{alg:Active_sync} ensures that the patches being synchronized enter the decoding stage at the same time. However, this does not mean that they will remain synchronized after decoding is complete, due to the non-deterministic decoding latency of every patch. To remedy this, the decoding latency for both patches can be fixed (at the cost of a possible temporary increase in the number of errors) or the use of \textit{Passive} synchronization where the patch decoded first waits for the second patch. The progress being made in designing accurate and faster decoders~\cite{astrea, caune2023belief, chamberland2022techniques, Gicev2023, smith2023predecoder} makes the former a more attrActive solution for the final synchronization step. 

%%%%%%%% Add text about new figures
% {\color{red} need some text on experimental step mimics the active and passive faithfully }

\subheading{Insights from experiments on IBMQ}
\textit{Active} synchronization aims to enhance the mitigation provided by Dynamical Decoupling (DD) against idling errors by reducing  idle periods experienced by data qubits. To demonstrate its effectiveness, we conducted two experiments on IBM Brisbane~\cite{ibm_brisbane}, a 127-qubit quantum computer. In the first experiment (Figure~\ref{subfig:ibm_Passive}), an idle period of $t_p$ was added after $N$ repetitions of a gate sequence, mimicking the \textit{Passive} policy. In the second experiment (Figure~\ref{subfig:ibm_Active}), the same idle period was split equally across all $N$ repetitions, with $t_a = t_p / N$, and added to the end of each gate sequence, emulating the \textit{Active} policy. Both circuits had the same total duration and were applied to the first 20 qubits of the system, with X-X DD sequences inserted during all idle periods.

The benefits of dividing idle durations are clear in Figure~\ref{subfig:ibm_charac_ap}, which shows the mean fidelity across all 20 qubits for $N=20$ and $N=200$. The circuit using the \textit{Active} policy consistently outperformed the \textit{Passive} policy. Moreover, as $N$ increases, the idle period $t_a$ becomes smaller, further improving performance compared to the \textit{Passive} policy. While the improvement with the \textit{Active} policy is modest for $t_p = 0.8\mu s$ at both $N=20$ and $N=200$, the benefits are expected to be more significant in error correction circuits involving more entangled qubits.

\begin{figure}[t]
    \centering
    % \hspace*{\fill}
    \begin{subfigure}[b]{0.37\linewidth}
        \centering
        \includegraphics[width=\textwidth]{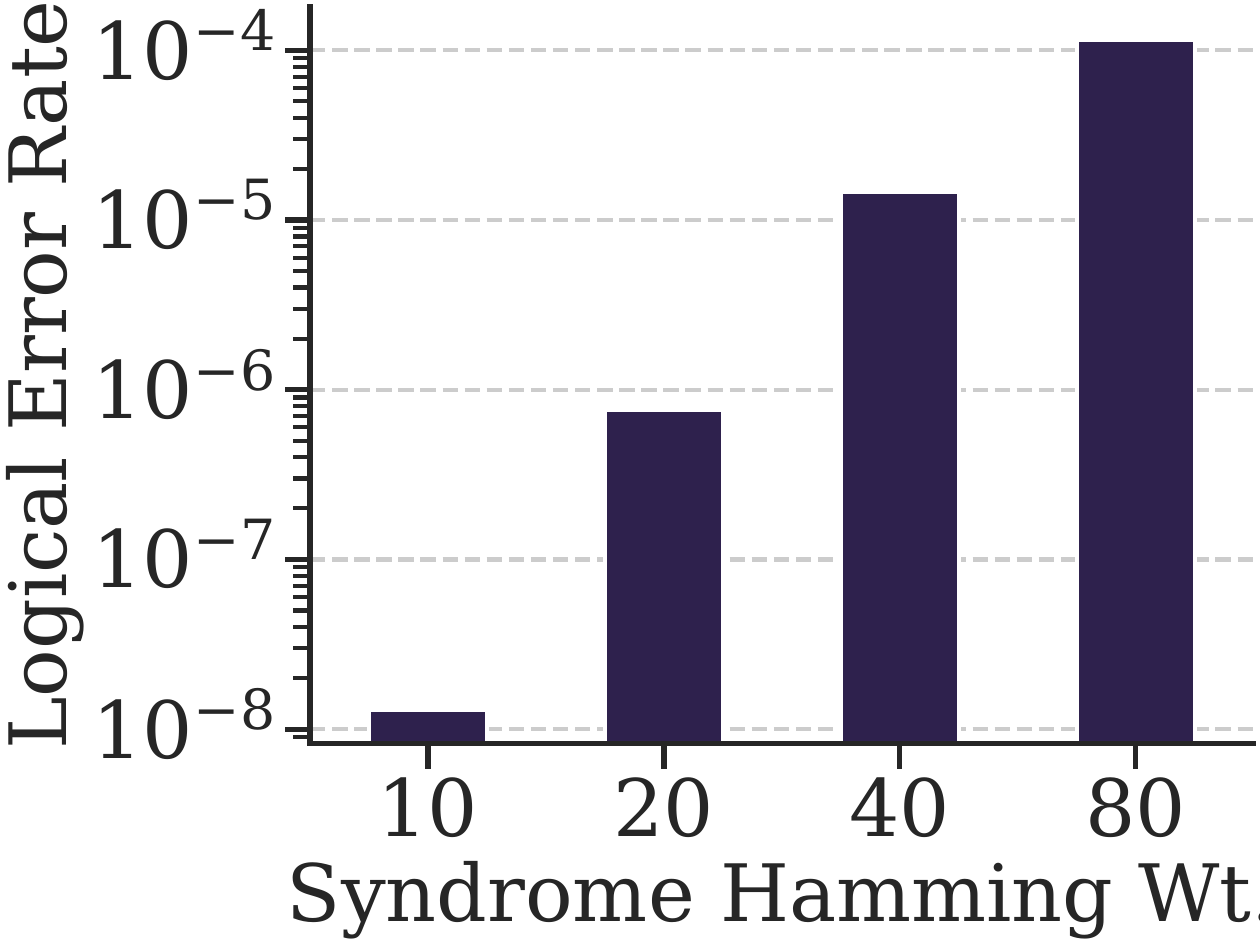}
        \caption{}
        \label{subfig:ler_hamm}
    \end{subfigure}
    \hspace*{\fill}
    \begin{subfigure}[b]{0.61\linewidth}
        \centering
        \includegraphics[width=\textwidth]{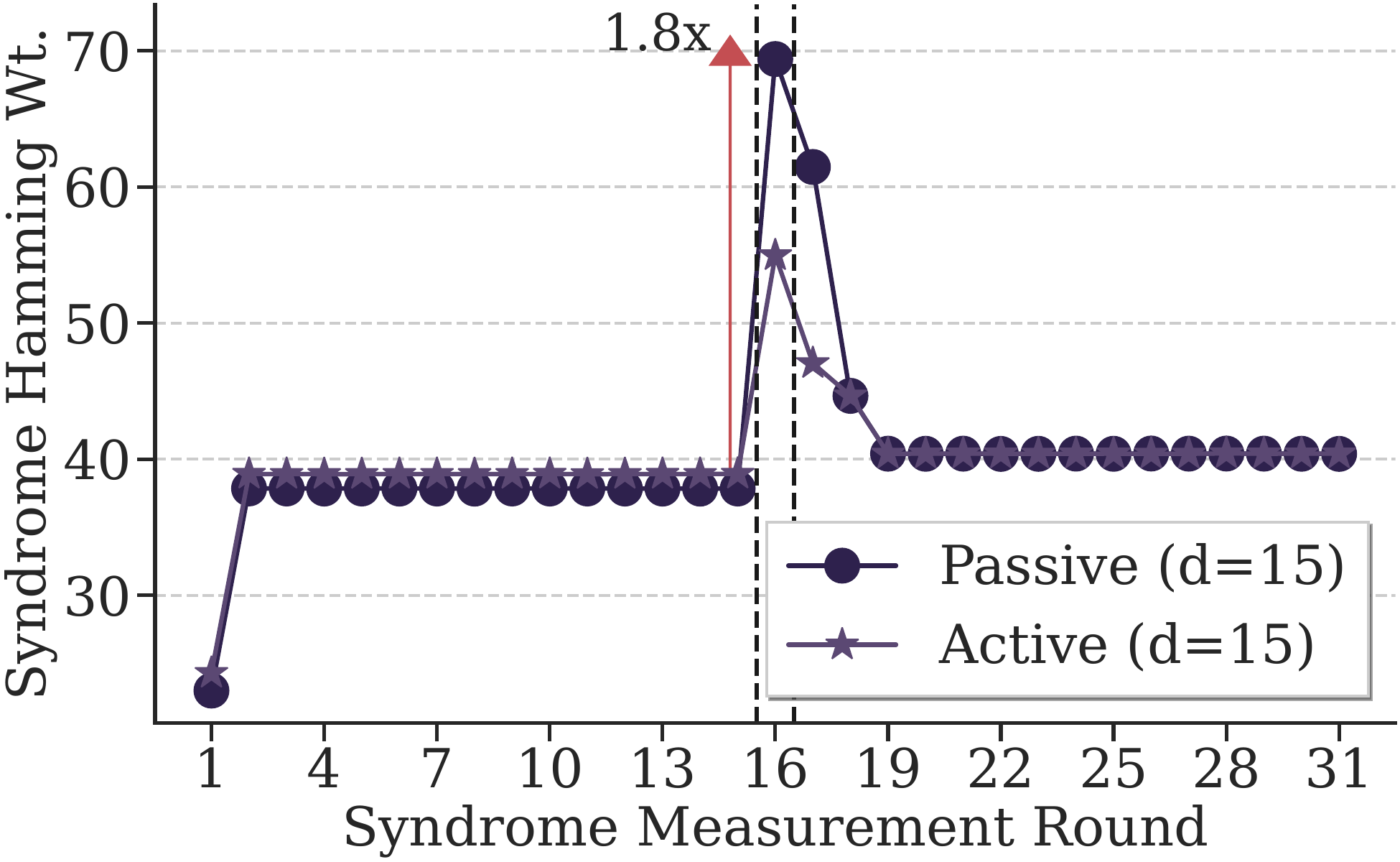}
        \caption{}
        \label{subfig:syn_hamm}
    \end{subfigure}
    % \hspace*{\fill}
    \caption{
    (a) Logical error rate is influenced by the syndrome hamming weight -- higher weights result in more logical errors ($d=15, p=0.001$);
    (b) Syndrome hamming weights for all rounds before and after Lattice Surgery with \textit{Passive} and \textit{Active} synchronization ($\tau=500$ns). The dashed vertical line shows the round during which Lattice Surgery is performed. \textit{Passive} synchronization results in an almost 2x higher hamming weight during Lattice Surgery, resulting in a higher chance of logical errors compared to \textit{Active} synchronization.
    }
    \vspace{-0.05in}
    \Description[Hamm wt.]{}
    \label{fig:hamm_syn_ler}
\end{figure}

% {\color{red}- refer to fig6a}

\subheading{Why does the \textit{Active} policy work?}

The experiments on IBM hardware highlighted the benefits of the \textit{Active} policy on physical qubits. To evaluate its impact on logical qubits, we analyzed syndromes collected during Lattice Surgery operations under both \textit{Active} and \textit{Passive} policies. As shown in Figure~\ref{subfig:ler_hamm}, logical error rates increase with higher syndrome Hamming weights, as these are more likely to introduce uncorrectable errors. Figure~\ref{subfig:syn_hamm} illustrates the syndrome Hamming weight\footnote{These syndromes correspond to detectors~\cite{Gidney2021}.} over QEC rounds for $d=15$. During Lattice Surgery, there is a spike in Hamming weight, which is significantly higher under the \textit{Passive} policy. Although the \textit{Active} policy results in a slightly higher average Hamming weight before Lattice Surgery, the spike during the operation is much smaller, leading to better logical error rates. This improvement can be seen in Table~\ref{tab:Active_example}, where the \textit{Active} policy reduces undetectable errors across various code distances ($p=10^{-3}$, $T_1=25\mu$s, $T_2=40\mu$s, $10^5$ shots).

% The previous experiment focused on the benefits of the \textit{Active} policy on just physical qubits. To understand its effect on a logical qubit, we sample syndromes received during the Lattice Surgery operation for both \textit{Active} and \textit{Passive} policies. As shown in Figure~\ref{subfig:ler_hamm}, the logical error rate degrades with higher syndrome hamming weights. This is because higher syndrome hamming weights are more likely to introduce uncorrectable errors. The syndrome hamming weight\footnote{Note that these syndromes are actually detectors~\cite{Gidney2021}.} as a function of QEC rounds is shown in Figure~\ref{subfig:syn_hamm} for $d=15$. In the round Lattice Surgery is performed, there is a spike in the hamming weight, and it is significantly higher for the \textit{Passive} policy. Thus, even though the \textit{Active} policy results in a slightly higher average hamming weight prior to Lattice Surgery, the increase during Lattice Surgery is significantly smaller than the \textit{Passive} policy, resulting in better error rates. This can be seen in Table~\ref{tab:Active_example}, where we see the \textit{Active} policy reduces the number of undetectable errors for different code distances ($p=10^{-3}$, $T_1=25\mu$s, $T_2=40\mu$s, $10^5$ shots).

\begin{table}[ht!]
\centering
\caption{Number of logical errors for different code distances $d$ for the example shown in Figure~\ref{fig:Active_Passive_sync}.}
\label{tab:Active_example}
\resizebox{0.95\linewidth}{!}{%
\begin{tabular}{ccccccccc}
\specialrule{1.5pt}{1pt}{1pt}
\multirow{2}{*}{\textbf{Case}}            & \multicolumn{4}{c}{Number of Errors (slack=500ns)}                                        & \multicolumn{4}{c}{Number of Errors (slack=1000ns)}                                                             \\ \cline{2-9} 
                                 & $d=7$                & $d=9$                & $d=11$               & $d=13$               & \multicolumn{1}{c}{$d=7$} & \multicolumn{1}{c}{$d=9$} & \multicolumn{1}{c}{$d=11$} & \multicolumn{1}{c}{$d=13$} \\ \hline
\textit{Passive}                          & 278                  & 89                   & 22                   & 5                    & 403                       & 114                       & 31                         & 13                         \\ 
\textit{Active}                           & 232                  & 77                   & 15                   & 2                    & 324                       & 94                        & 18                         & 6                          \\ 
\multicolumn{1}{l}{\% Reduction} & \multicolumn{1}{c}{16.54} & \multicolumn{1}{c}{13.48} & \multicolumn{1}{c}{31.81} & \multicolumn{1}{c}{60} & 19.6                      & 17.54                     & 41.93                      & 53.84                      \\ \specialrule{1.5pt}{1pt}{1pt}
\end{tabular}
}
\end{table}

% \begin{figure}[htbp]
% \centering
% \begin{quantikz}[row sep={0.7cm,between origins}, column sep =3mm]
%   & \lstick{$\ket{D}$}& \ctrl{1} & \ctrl{2} & \ctrl{3} & \ctrl{4} & \qw & \qw & \qw & \qw & \qw & \qw & \qw & \qw & \qw & \qw \\
%   & \lstick{$\ket{0}$} & \targ{}  & \qw & \qw & \qw & \meter{} & &\lstick{$\ket{0}$} & \qw & \qw \\
%   & \lstick{$\ket{0}$} & \qw &  \targ{} & \qw & \qw & \meter{} & &\lstick{$\ket{0}$} & \qw & \qw \\
%   & \lstick{$\ket{0}$} & \qw & \qw & \targ{} & \qw & \meter{} & &\lstick{$\ket{0}$} & \qw & \qw \\
%   & \lstick{$\ket{0}$} & \qw & \qw & \qw & \targ{} & \meter{} & &\lstick{$\ket{0}$} & \qw & \qw \\
% \end{quantikz}
% \caption{Quantum circuit with initialized qubits and CNOT gates, leading to measurements.}
% \label{fig:quantum_circuit}
% \end{figure}
\subsubsection{\textit{Active} synchronization within a single round}

\begin{figure}[t]
    \centering
    \includegraphics[width=0.99\linewidth]{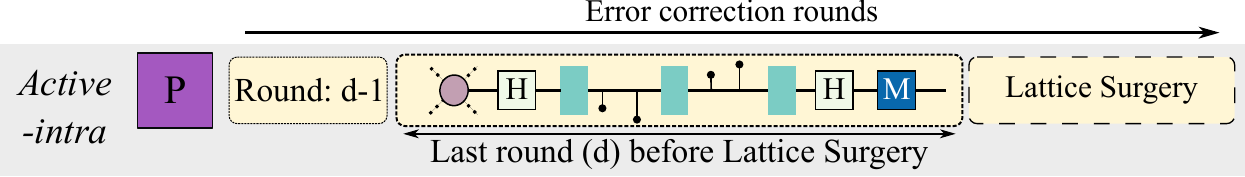}
    \caption{The \textit{Active-intra} policy, where the slack is distributed within the last syndrome measurement round before Lattice Surgery.}
    \vspace{-0.1in}
    \label{fig:Active_intra}
    \Description[ActiveIntra]{}
\end{figure}

Another variant of the \textit{Active} policy is to distribute the total slack within a round rather than between multiple rounds. We call this the \textit{Active-intra} policy. This can be seen in Figure~\ref{fig:Active_intra} where the slack is distributed inside the final round before Lattice Surgery. An advantage of this policy is that synchronization can be performed within one round rather than multiple rounds. An obvious disadvantage is that this would add decoherence errors due to idling on the measure qubits as well as the data qubits. We will show in Section~\ref{sec:eval} that the \textit{Active-intra} policy is generally inferior to the \textit{Active} policy due to this.

\subsubsection{Synchronizing by running additional rounds}
\begin{figure}[t]
    \centering
    \includegraphics[width=0.99\linewidth]{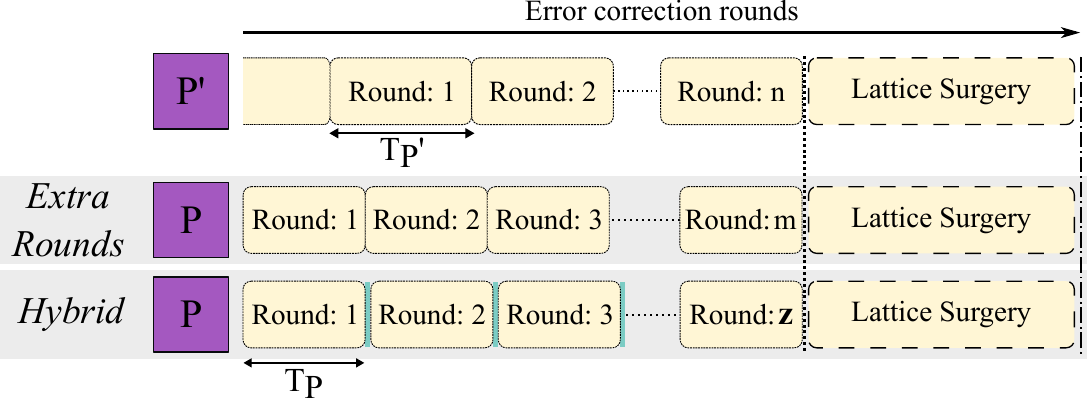}
    \caption{
    Extra Rounds: for patch $P'$ with a cycle time of $T_P'$ and $P$ with a cycle time of $T_P$, synchronization before Lattice Surgery can be achieved by running $P(P')$ for $m(n)$ extra rounds; \textit{Hybrid} policy: a small idling period is added after every round and $P$ is run for $z$ extra rounds.}
    \vspace{-0.1in}
    \label{fig:additional_rounds_des}
    \Description[addRounds]{}
\end{figure}

Rather than stopping syndrome generation for the leading patch $P$ and idling for the slack duration, logical qubits can also be synchronized by running additional rounds. \textbf{This is not possible if both $P$ and $P'$ have the same syndrome generation cycle time ($T_P' = T_P$)}. Figure~\ref{fig:additional_rounds_des} shows how patch $P'$ with a cycle time of $T_P'$ can be run for $n$ rounds, while patch $P$ with a cycle time of $T_P$ can be run for $m$ rounds to synchronize the two patches. In general, this synchronization policy can be defined with a Diophantine equation as: 

% \vspace{-0.15in}
\begin{equation}
\label{eqn:extra_rounds}
    n.T_P' = m.T_P + \tau;~~~(T_P \neq T_P')
\end{equation}

\begin{figure}[t]
    \centering
    \includegraphics[width=0.75\linewidth]{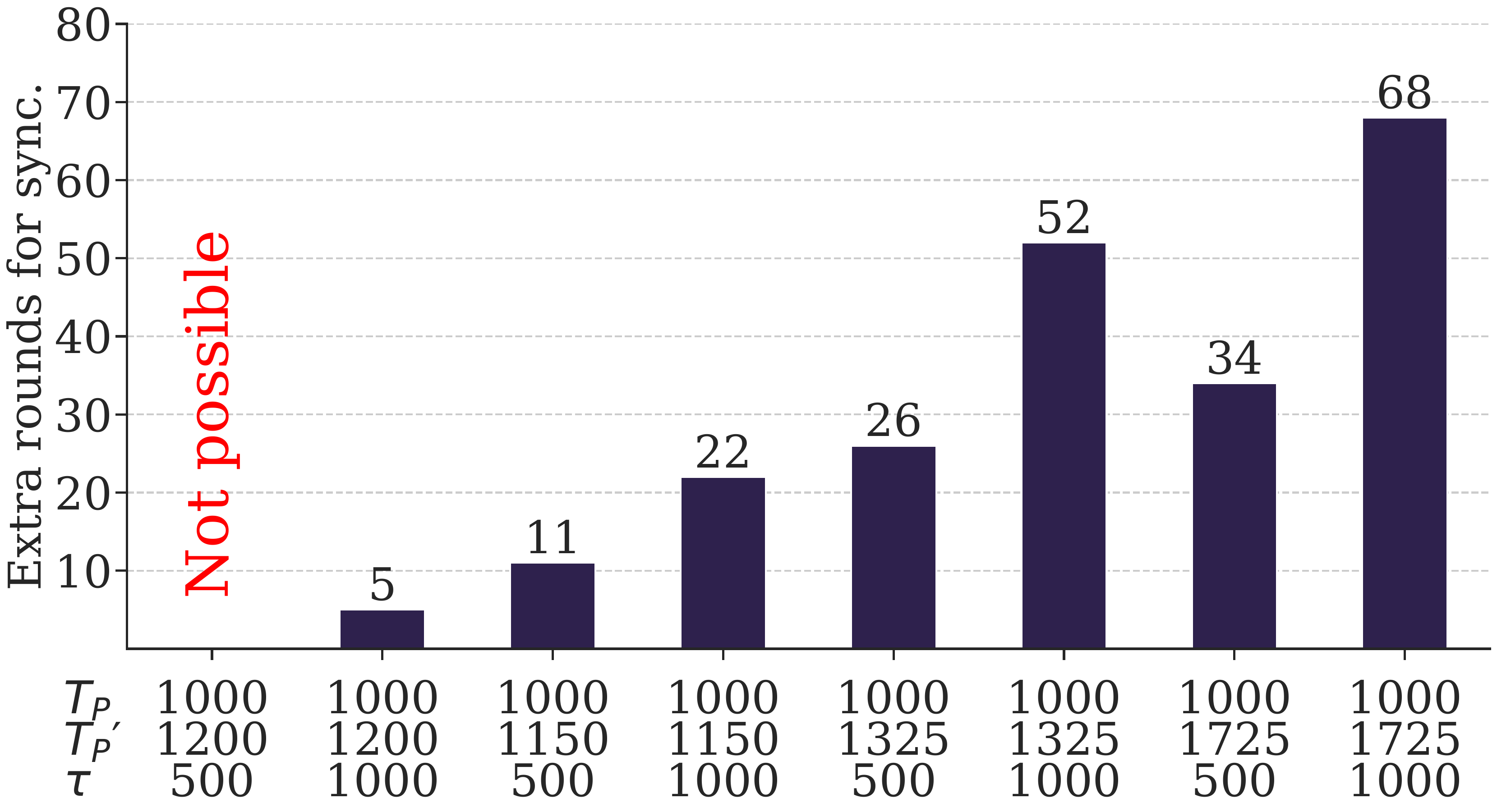}
    \caption{
    \rev{
    Different example configurations to show how running $P$ for $m$ extra rounds to achieve synchronization can be advantageous in some cases and impossible in others with the syndrome generation cycle times of $P, P'$ as $T_{P},T_{P}'$ (ns) and an initial slack $\tau$ (ns).
    }
    }
    \vspace{-0.1in}
    \label{fig:extra_rounds}
    \Description[extraRounds]{}
\end{figure}

\noindent Here $\tau$ is the synchronization slack. To synchronize the two patches, we must iteratively find the smallest integer $m$ such that $\frac{m.T_P + \tau}{T_P'}$ is an integer. As shown in Figure~\ref{fig:extra_rounds}, the number of additional rounds required is highly dependent on the values of $\tau, T_P', T_P$, and in some cases, has no integral solution. As is evident from Figure~\ref{fig:extra_rounds}, some combinations of $\tau, T_P', T_P$ can require numerous rounds \textbf{per synchronization}, which will cause significant slowdown and reduce the per round error budget for other logical operations.

\subsection{Combining Policies}
It is impossible to synchronize patches $P$ and $P'$ by just running additional rounds if their syndrome generation cycle times are equal ($T_P = T_P'$). Thus, for $T_P = T_P'$, they will have to be synchronized by using either \textit{Active} or \textit{Passive} synchronization. However, when $T_P \ne T_P'$, there is an opportunity to further optimize the synchronization policy by running some additional rounds of error correction \textbf{and} using the \textit{Active} policy. This can be visualized in Figure~\ref{fig:additional_rounds_des}, where the \textit{Hybrid} policy inserts significantly smaller idling periods between the rounds while running for an additional $z$ rounds. By combining the two policies, we can reduce the total idle period during which syndrome generation is paused compared to the pure \textit{Active} policy implementation and also reduce the number of rounds required compared to the pure extra round strategy. To find an optimal \textit{Hybrid} policy for patches $P$, and $P'$, we first define a slack tolerance, $\epsilon$, which is the maximum duration for which syndrome generation can be paused. Using $\epsilon$, the value of $z$ can be computed by finding the smallest integer $z$ for which --

% \vspace{-0.1in}
\begin{equation}
\label{eqn:hybrid}
    \ceil*{\frac{z.T_P + \tau}{T_P'}} \times T_P' - (z.T_P + \tau) < \epsilon;~~~(T_P \neq T_P')
\end{equation}

\begin{table}[ht!]
\centering
\vspace{-0.05in}
\caption{
\rev{Comparison of the idling period, extra rounds, and LER of different synchronization policies for the given configuration of patch cycle times and initial slack $\tau$. The \textit{Hybrid} policy achieves a significantly better LER by reducing the idling period and the number of extra rounds run.}
}
\label{tab:hybrid}
\resizebox{0.85\linewidth}{!}{%
\begin{tabular}{c|ccc} 
\specialrule{1.5pt}{1pt}{1pt}
\multicolumn{4}{c}{$\mathbf{T_P=1000ns, T_P'=1325ns, {\tau=1000ns}, {\epsilon=400ns}}$, 20M shots}                 \\ \specialrule{1.25pt}{1pt}{1pt} \specialrule{1.25pt}{1pt}{1pt}
\textbf{Synchronization Policy}       & \textit{Active} & \textit{Extra Rounds} & \textbf{\textit{Hybrid}} \\ \hline
\textbf{Idling period} & 1000 ns                  & 0                 & 300 ns \\ 
\textbf{Num. extra rounds} & 0                       & 52                & 4      \\ 
\textbf{LER ($d=7$})    & 0.0014                        &  0.0059                 & 0.00095       \\ \specialrule{1.5pt}{1pt}{1pt}
\end{tabular}
}
\end{table}

As an example, we show how the three policies compare for a specific scenario where $t_P=1000ns$, $t_P'=1325ns$, $\tau=800ns$, $\epsilon = 200ns$. Here, the \textit{Hybrid} synchronization policy can reduce the idling duration to 175ns from 800ns and the number of rounds from 31 to 3. This combination results in a logical error rate (LER) reduction of 1.71x compared to the pure \textit{Active} policy and a 3.25x reduction compared to when only additional rounds are run. While this is a very specific case where the \textit{Hybrid} policy works much better, there are cases where the \textit{Hybrid} policy cannot do better than just running extra rounds or the \textit{Active} policy. We report results for many other cases in Section~\ref{sec:eval}.

\begin{figure}[t]
    \centering
    \includegraphics[width=0.99\linewidth]{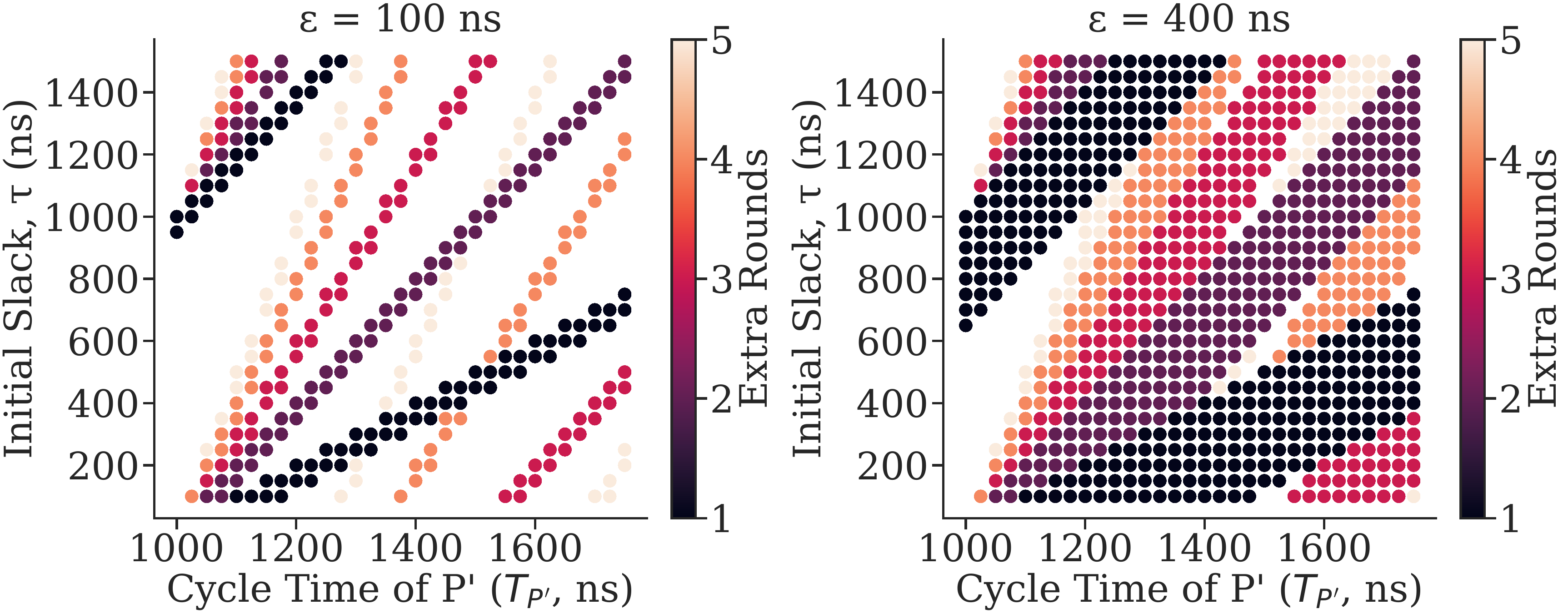}
    \caption{
    \rev{
    $\tau$ vs. $T_{P'}$, colored by the number of extra rounds ($z$). $T_P$ was fixed at 1000ns. $\epsilon=400ns$ allows more $\tau, T_P/T_{P'}$ combinations to be optimized with the \textit{Hybrid} policy. White space indicates no solution was found. 
    }
    }
    \vspace{-0.1in}
    \label{fig:hybrid_opt}
    \Description[addRounds]{}
\end{figure}

\rev{
\subsubsection{Optimizing the \textit{Hybrid} policy}
\label{sec:hybrid_optimize}
A natural question that arises from Equation~\ref{eqn:hybrid} is the selection of the slack tolerance ($\epsilon$). As with the Diophantine equation defined in Equation~\ref{eqn:extra_rounds}, Equation~\ref{eqn:hybrid} can only be solved iteratively, thus requiring some bounds to be set on $z$ and $\epsilon$. For this work, we set an upper bound on the number of extra rounds and the maximum tolerated slack with the \textit{Hybrid} policy at 5 rounds and 400ns respectively. Note that this will depend on the noise profile of the system and the cycle times $T_P, T_{P'}$ of the patches being synchronized. Figure~\ref{fig:hybrid_opt} shows the trade-off between using a smaller value of $\epsilon = 100ns$ and a larger one $\epsilon=400ns$. Using a smaller value of $\epsilon$ will thus be benefical if there is a very limited variance in the slack ($\tau$) and the cycle times ($T_P, T_{P'}$). Using larger values of $\epsilon$ allows for more combinations of $\tau, T_P, T_{P'}$ to be tackled with the \textit{Hybrid} policy. Due to its more general application, we use a larger value of $\epsilon=400ns$ for all evaluations.
}

\subsection{Synchronizing \textit{k} Patches}
\label{sec:syncK}

Generalizing two patch synchronization to an arbitrary number of patches can be done by identifying the slowest (most lagging) patch after sorting the patches based on the time needed for them to complete their current code cycle and then performing pair-wise synchronizations of all the patches with the slowest patch. All pair-wise synchronizations can be performed in parallel, which makes the synchronization of $k$ patches take constant time. 

\section{Microarchitecture for Synchronization}
\label{sec:uarch}

% In a large-scale system with multiple controllers responsible for all qubits, there can be a skew between the controllers themselves. Assuming that every controller has access to a master clock, the start of every synchronization must begin with a synchronization of all the hardware controllers. Once the controllers are synchronized, the elapsed time in the current code cycle for all patches can be determined accurately. 
{Synchronization is performed at {\em runtime}. Determining the synchronization slack between two or more logical patches requires the knowledge of (i) the cycle duration of each patch ($T$), and (ii) the time elapsed in the current code cycle for each patch ($t$). The cycle duration depends on the gate and measurement latencies of the stabilizer circuit, which in turn are dependent on the physical qubits that constitute a patch. 
% Since Lattice Surgery modifies the shape and position of patches during computation, the cycle duration of a patch can keep changing. 
Assuming that the physical qubits in the lattice are calibrated regularly, the cycle duration of all patches required for a computation (in time and space) can be computed and stored in a table during compile time. This is possible since Lattice Surgery compilers can generate an Intermediate Representation (IR) of all Lattice Surgery operations required to implement a circuit~\cite{watkins2023high, Leblond2023}.  Patches are merged/split to create new patches, the entire IR can be used to map physical qubits to their corresponding logical patches, from which the cycle duration can be determined. }

{
Determining the time elapsed in the current code cycle ($t$) for every patch that needs to be synchronized will be a more involved process. To know the current phase of every logical patch in the system will require the control hardware to maintain counters for every logical patch in the system. Each counter will increment at every tick of the global clock, and will start and end with the surface code cycle for its corresponding patch. Once a patch has been merged/split to yield a different patch(s), the counter for the former patch can be disabled. As surface code cycles for superconducting qubits can lie between 1000–2000~ns~\cite{GoogleSuppressing, Sundaresan2023} and assuming a clock frequency of 1~GHz, 10-12 bit counters per logical patch will be enough to determine $t$.
}

\begin{figure}[t]
    \centering
    \includegraphics[width=0.9\linewidth]{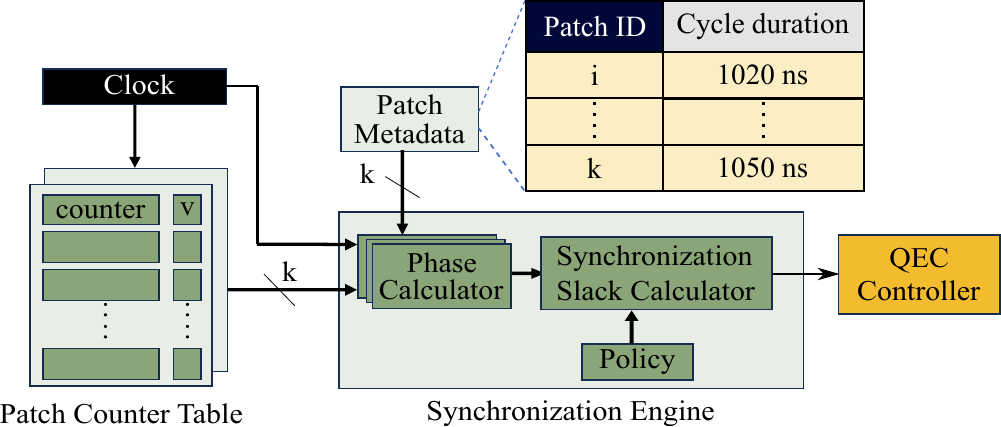}
    
    \caption{Microarchitecture for enabling synchronization.}
    \vspace{-0.15in}
    \Description[a figure]{}
    \label{fig:uarch}
    
\end{figure}

Figure~\ref{fig:uarch} shows a control microarchitecture for enabling synchronization -- the Patch Counter Table contains information about the number of rounds completed for every patch, and the valid bit specifies whether the patch is a valid patch or not. The information from the patch metadata table -- which contains information on the cycle duration of every patch -- and the patch counter table of the $k$ patches that need to be synchronized is passed to the synchronization engine which determines the synchronization slack that needs to be added to the schedule. This is done by first determining the difference in execution phase of the patches through the phase calculator to determine the fastest and slowest patch. Once this information is known, the slack calculator can insert the appropriate barriers in the schedule of every patch involved to synchronize them. 
In case the \textit{Hybrid} policy is used, the synchronization engine can solve Equation~\ref{eqn:hybrid} using the slack and the cycle durations to determine the number of additional rounds to run and the new slack that must be inserted for synchronization. The synchronized schedule is then passed to the controller for execution. 
% The patch metadata must be updated after every Lattice Surgery operation, since the cycle duration of a patch will change after every split and merge operation. 

Choosing the appropriate synchronization policy can be performed at runtime too. Since the \textit{Hybrid} policy requires a slack tolerance $\epsilon$, it will only give a valid combination of slack time and additional rounds if the slack is less than $\epsilon$. To select between the \textit{Active} and \textit{Hybrid} policy thereafter, a simple decision can be made based on calibration data to determine which slack and round combinations suit the \textit{Hybrid} policy more.

While synchronization will need to be performed at runtime, the synchronization engine falls outside the critical path as far as latency is concerned since every logical operation is performed at a cadence of $d$ rounds of error correction at a minimum~\cite{Litinski2019, fowler2019low}. Thus, determining the current execution phase of the logical qubits involved in the next logical operation can be performed within the current $d$ code cycles (in the order of a few microseconds), which is enough time for the engine to calculate the slack and define a synchronized schedule for all the logical qubits involved.
\section{Methodology}

% \begin{figure}[t]
%     \centering
%     \includegraphics[width=0.8\linewidth]{figs/growth.pdf}
%     \caption{Growth of the mean slack between logical qubits for \textit{qft-12} normalized to the code cycle time(=1000ns).}
%     \Description[a figure]{}
%     \label{fig:growth}
% \end{figure}

\begin{figure}[t]
    \centering
    \includegraphics[width=\linewidth]{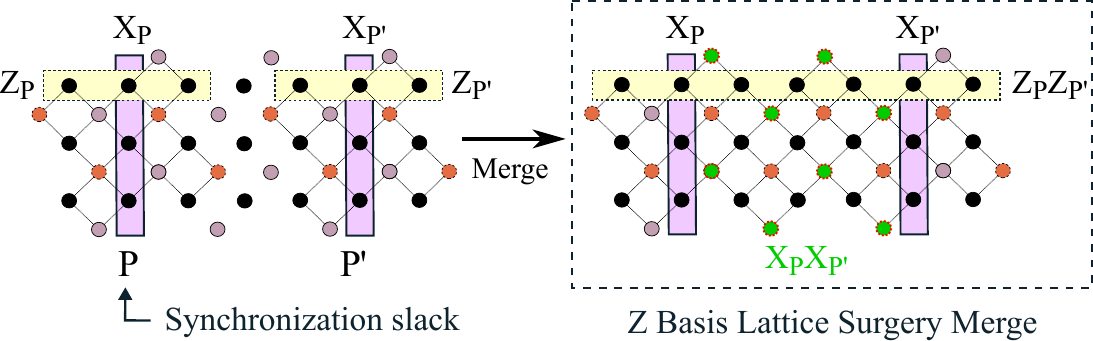}
    
    \caption{Two surface code patches $P$, $P'$ are initialized, with some synchronization slack added to $P$. After $d+1$ rounds, they are synchronized and then merged, after which the merged patch is run for another $d+1$ rounds. The observables $X_P$, $X_{P'}$, $Z_{P}Z_{P'}$, $X_{P}X_{P'}$ of the merged patch can be used by Stim to compute the Logical Error Rate (LER). 
    Note that for the $Z$ basis Lattice Surgery, the $X_{P}X_{P'}$ observable corresponds to the logical measurement between $P$ and $P'$ and is computed by measuring the green $X$ measure qubits~\cite{fowler2019low}.
    }
    \vspace{-0.1in}
    \Description[methodology]{}
    \label{fig:methodology}
    
\end{figure}

\subheading{Simulator}
The simulation framework used in this work is a combination of our proposed framework \texttt{lattice-sim} and Stim~\cite{Gidney2021}. \texttt{lattice-sim} generates stabilizer circuits compatible with Stim and can also perform Lattice Surgery between two patches, as shown in Figure~\ref{fig:methodology} (\texttt{lattice-sim} can perform both $X$ and $Z$ basis Lattice Surgery). We will focus on the $X_{P}X_{P'}$, $X_P$ ($Z_{P}Z_{P'}$, $Z_P$) observables for $Z$ ($X$) basis Lattice Surgery for the sake of brevity. While the use of heterogeneous codes will most likely be the dominant source of synchronization in FTQC systems, we restrict our evaluations to only surface code patches because decoders are faster and more accurate for the surface code compared to other codes such as the Color code. Building decoders that can accurately decode Lattice Surgery operations between different codes is an active field of research~\cite{shutty2022decoding}.
% The Logical Error Rate of the merged patch is computed only for the $X_{P}$ and $X_{P}X_{P'}$ observables ($Z_{P}$ and $Z_{P}Z_{P'}$ for $X$ basis Lattice Surgery). 
% We omit the remaining observables for the sake of brevity. In particular, we choose the $X_{P}X_{P'}$ observable for $Z$ basis and the $Z_{P}Z_{P'}$ observable for $X$ basis Lattice Surgery since they represent a logical measurement between two patches\footnote{Note that a logical measurement is followed by a split~\cite{fowler2019low}, which we do not simulate.}, which is the primary use case for Lattice Surgery~\cite{Litinski2019}. 
% \textbf{\texttt{lattice-sim} is the one of the first open-source and scalable Lattice Surgery stabilizer simulators}.

% As shown in Figure~\ref{fig:simulator}, 
\texttt{lattice-sim} consists of a parser that can take QASM~\cite{qasm} circuits as an input to generate the corresponding stabilizer circuits (non-Clifford operations are not supported, as Stim is a Clifford simulator). \texttt{lattice-sim} also has a rich error modeling interface to add error annotations to the stabilizer circuit for realistic and device-specific simulations \textbf{with circuit-level noise} of $p=10^{-3}$. \texttt{lattice-sim} also annotates idling errors based on the idle periods experienced by the qubits after every operation. The idling error model is based on the Pauli twirl approximation~\cite{Ghosh2012, Svore2014}.
% {that reduces the problem of characterizing a general quantum channel to that of characterizing a finite set of Pauli channels}. The relaxation ($T_1$) and decoherence ($T_2$) times can be fixed to a single value or can be sampled from a distribution. 
Idling errors were inserted as single Pauli error channels~\cite{Flammia2020}, with $p_x$, $p_y$, and $p_z$ defined as: $p_x=p_y=\frac{1-e^{-\tau/T_1}}{4}; p_z=\frac{1-e^{-\tau/T_2}}{2} - p_x$.

% \vspace{-0.05in}
% \begin{equation}l
% p_x=p_y=\frac{1-e^{-\tau/T_1}}{4}; p_z=\frac{1-e^{-\tau/T_2}}{2} - p_x
% \end{equation}

Note that the idling error model is conservative and does not model crosstalk, spectator effects, and leakage, which are typically observed on real hardware. 

% The scheduler in \texttt{lattice-sim} implements both \texttt{fenced} and \texttt{runahead} scheduling policies with support for both \textit{Active} and \textit{passive} synchronization mechanisms. \texttt{lattice-sim} can simulate an arbitrary number of patches and  can model skew between distributed qubit controllers. 

% Future development of \texttt{lattice-sim} will include the support of Lattice Surgery \texttt{split} and \texttt{merge} operations.

\subheading{Stim Configuration} 
\texttt{pymatching2}~\cite{Higgott2023} was used for decoding error syndromes. All circuits were sampled for 100M shots. The patches were initialized and run for $d+1$ rounds, after which they were merged using Lattice Surgery and run for another $d+1$ rounds.

\subheading{Hardware Configurations} 
Table~\ref{tab:configs} shows the hardware configurations used during simulations for evaluating different synchronization policies. The cycle time is the total time taken (gates + measurement + reset) for generating one round of syndromes.

\begin{table}[ht!]
\centering
\caption{System parameters used in evaluations.}
\label{tab:configs}
\resizebox{0.99\linewidth}{!}{%
\begin{tabular}{lccccc}
\specialrule{1.5pt}{1pt}{1pt}
\multicolumn{1}{c}{\textbf{Technology}} & $\mathbf{T_1, T_2}$ & \textbf{1Q gates} & \textbf{2Q gates} & \textbf{Readout} & \textbf{Cycle time $T_{cycle}$} \\ \hline
IBM~\cite{ibm_fez, ibm_brisbane}                                     & 200$\mu s$,150$\mu s$     & 50ns        & 70ns        & 1500ns           & $\sim$1900ns        \\ 
Google~\cite{acharya2024quantumerrorcorrectionsurface}                                  & 25$\mu s$,40$\mu s$       & 35ns        & 42ns        & 660ns            & $\sim$1100ns        \\ 
QuEra~\cite{Bluvstein2023, Marton2024}                                   & 4s,1.5s       & 5$\mu s$         & 200$\mu s$       & 1ms              & $\sim$2ms           \\ \specialrule{1.5pt}{1pt}{1pt}
\end{tabular}
}
\end{table}

\subheading{Other Software}
Benchmarks were generated and derived from MQTBench~\cite{quetschlich2023mqtbench}. Resource estimation was done using the Azure Quantum Resource Estimation tool~\cite{beverland2022assessing}.

% \begin{figure}[t]
%     \centering
%     \includegraphics[width=\linewidth]{figs/simulator.pdf}
%     \caption{\texttt{lattice-sim} organization -- Stim is used for simulating the stabilizer circuits generated by \texttt{lattice-sim}.}
%     
% \end{figure}

% \begin{figure}[t]
%     \centering
%     \hspace*{\fill}
%     \begin{subfigure}[b]{0.45\linewidth}
%         \centering
%         \includegraphics[width=\textwidth]{figs/color_surface.pdf}
%         \caption{}
%         \label{subfig:color-surface}
%     \end{subfigure}
%     \hspace*{\fill}
%     \begin{subfigure}[b]{0.45\linewidth}
%         \centering
%         \includegraphics[width=\textwidth]{figs/active_passive_color.pdf}
%         \caption{}
%         \label{subfig:color_active}
%     \end{subfigure}
%     \hspace*{\fill}
%     \caption{
%     \rev{
%     (a) Lattice Surgery between a color code and surface code patch~\cite{ito2024};
%     (b) Reduction in LER with the use of \textit{Active} synchronization compared to \textit{Passive}.
%     }
%     }
%     \vspace{-0.1in}
%     \Description[intra]{}
%     \label{fig:color}
% \end{figure}

\begin{figure*}[t]
    \centering
    \begin{subfigure}[b]{0.49\linewidth}
        \centering
        \includegraphics[width=\linewidth]{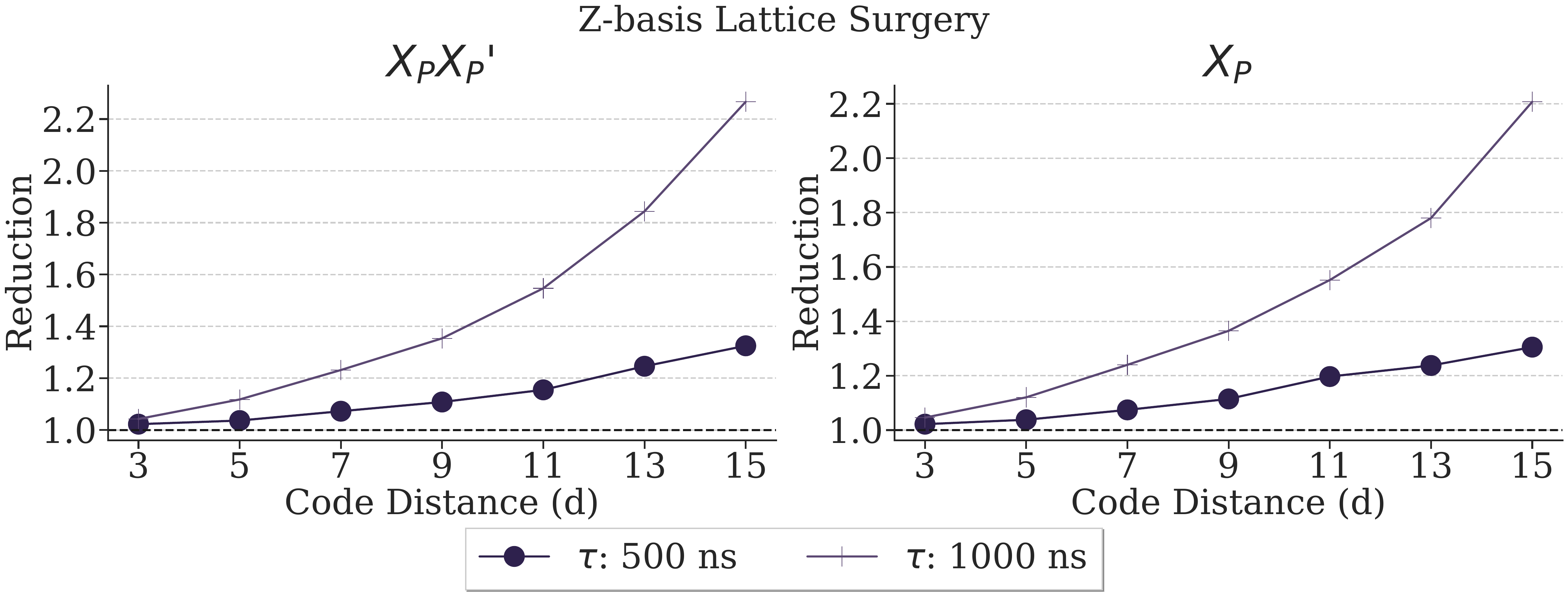}
        \caption{}
        \label{fig:ibmZ}
    \end{subfigure}
    \hfill
    \begin{subfigure}[b]{0.49\linewidth}
        \centering
        \includegraphics[width=\linewidth]{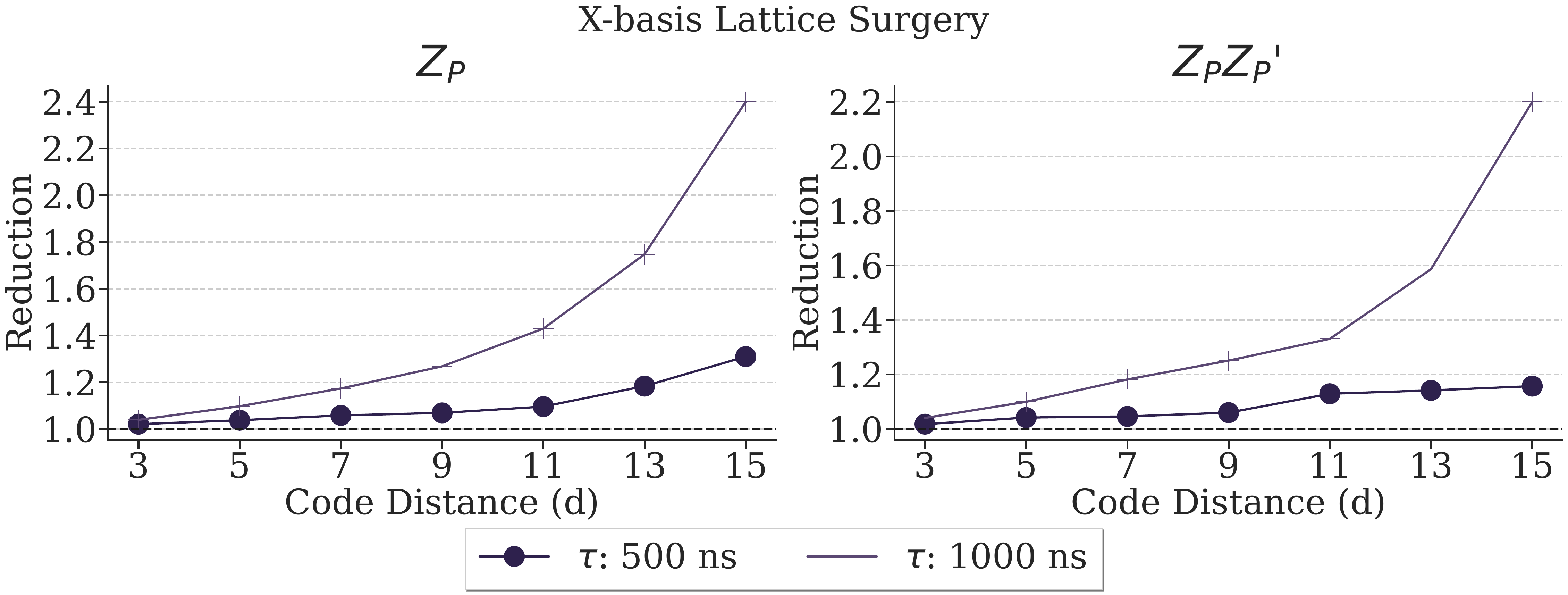}
        \caption{}
        \label{fig:ibmX}
    \end{subfigure}
    \\
    \begin{subfigure}[b]{0.49\linewidth}
        \centering
        \includegraphics[width=\linewidth]{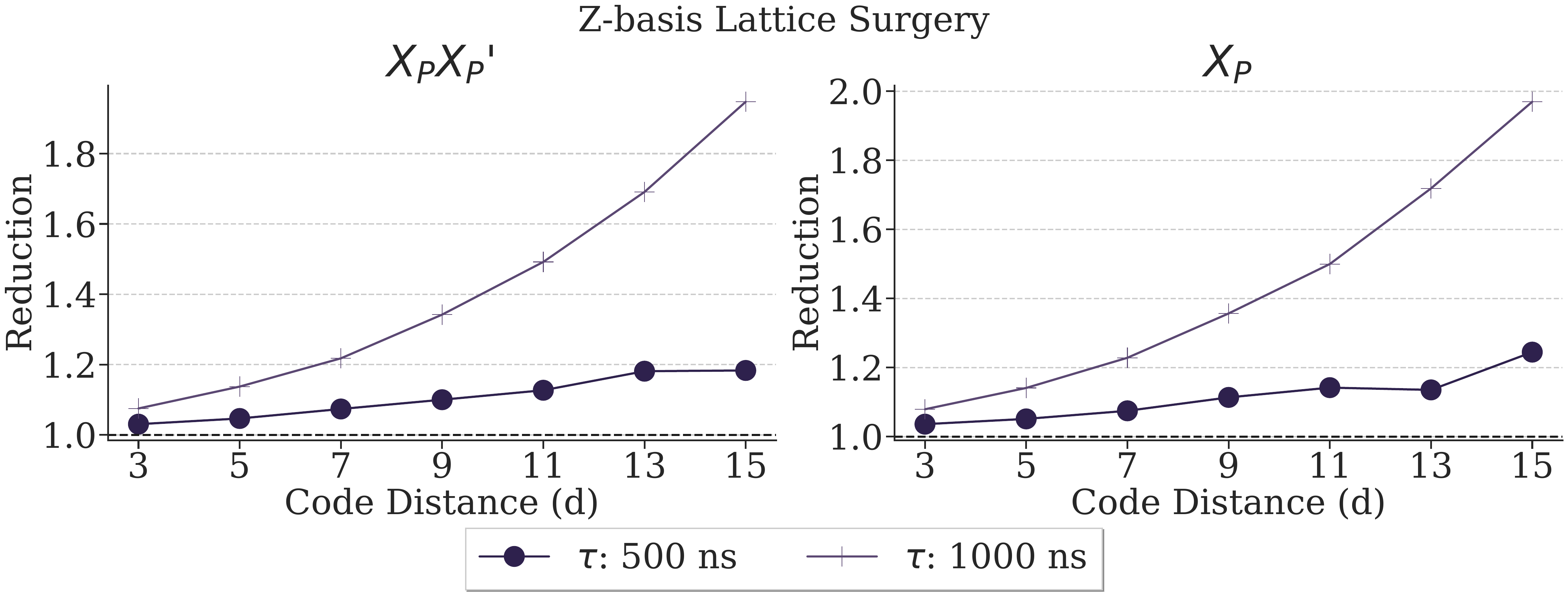}
        \caption{}
        \label{fig:googleZ}
    \end{subfigure}
    \hfill
    \begin{subfigure}[b]{0.49\linewidth}
        \centering
        \includegraphics[width=\linewidth]{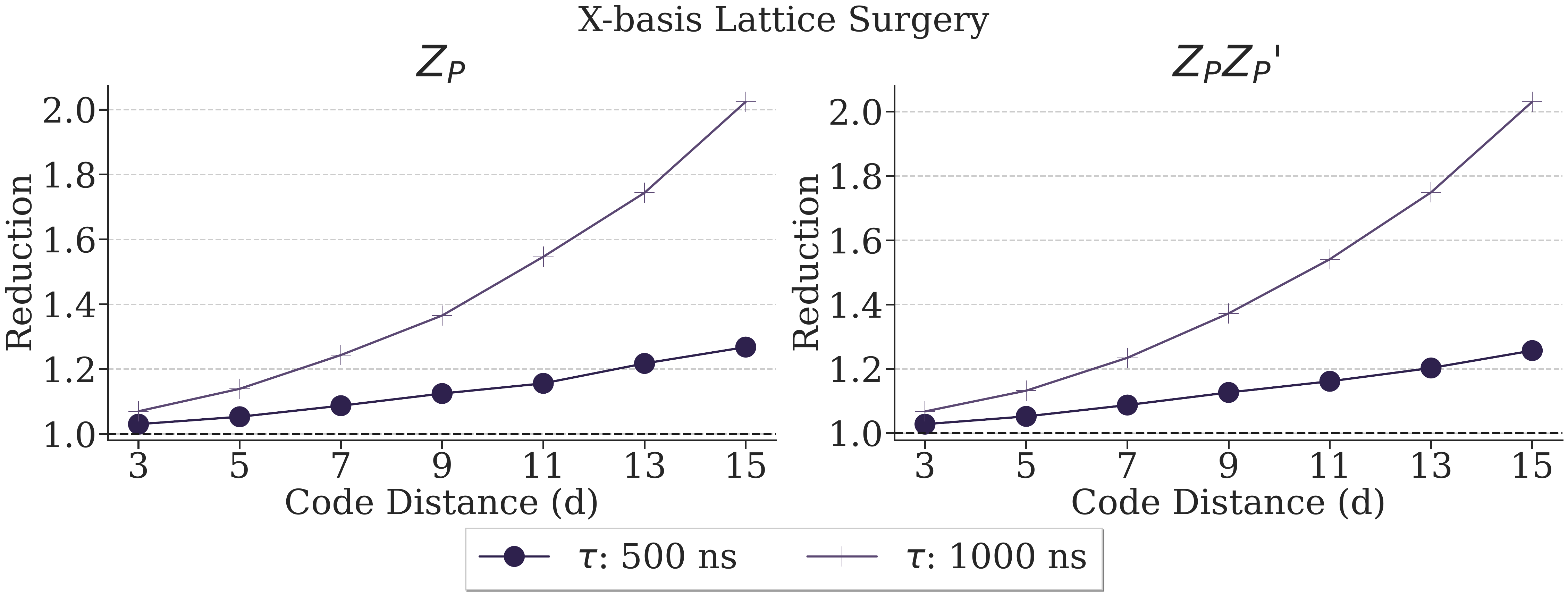}
        \caption{}
        \label{fig:googleX}
    \end{subfigure}
    
    \caption{
    \rev{Reduction in the logical error rate for \textbf{one synchronization} when using \textit{Active} synchronization instead of \textit{Passive} for different code distances with a system configuration similar to (a)-(b) IBM systems, and (c)-(d) Google systems for both $X$ and $Z$ basis Lattice Surgery with $p=10^{-3}$. The synchronization slack ($\tau$) was set at 500ns and 1000ns.}
    }
    \vspace{-0.1in}
    \Description[main results]{}
    \label{fig:sync_results}
    
\end{figure*}

\section{Evaluations}
\label{sec:eval}

% \begin{figure*}[hbtp]
%     \centering
%     \includegraphics[width=\linewidth]{figs/improvements.pdf}
%     \caption{Relative improvement in the logical error rate when using \textit{Active} synchronization instead of \textit{Passive} for different code distances with a system configuration similar to (a) IBM systems and (b) Google systems. The synchronization slack ($\tau$) has been varied from 100ns to 1000ns. Both \textit{Active} and \textit{Passive} synchronizations were applied to patches with \texttt{fenced} (left) and \texttt{runahead} (right) scheduling.}
%     \label{fig:sync_results0}
%     
% \end{figure*}

% Having described \textit{Active} synchronization and \texttt{runahead} scheduling, we will discuss evaluation results to quantify the efficacy of these policies. 
% To evaluate the proposed synchronization policy for the surface code, we developed a simulation framework \texttt{lattice-sim} that can synthesize stabilizer circuits with noise and latency annotations based on the synchronization policy. Figure~\ref{fig:methodology} shows how \texttt{lattice-sim} can generate two patches of an arbitrary code distance $d$, synchronize them depending on the slack configured, and then perform Lattice Surgery to merge them. All stabilizer circuits were simulated using Stim~\cite{Gidney2021}. \textbf{\texttt{lattice-sim} is the one of the first open-source and scalable Lattice Surgery simulators}. Using \texttt{lattice-sim}, 

\subsection{Research Questions}
We aim to answer the following key questions:

\begin{itemize}[topsep=0pt, leftmargin=*]
    \item Does \textit{Active} sync. reduce the impact of idling errors?
    \item What is the gap between an ideal hypothetical system that does not require synchronization?
    \item How do the \textit{Extra Rounds} and \textit{Hybrid} policies compare with the \textit{Active} and \textit{Passive} policies? 
    \item Which synchronization policies suit superconducting and neutral atom systems? 
    \item How does the synchronization policy affect performance?
    % \item How do defects in the lattice affect the efficacy of \textit{Active} synchronization?
\end{itemize}

% \begin{itemize}
%     \item Does \textit{Active} sync. reduce the impact of idling errors?
%     \item What is the gap between an ideal hypothetical system that does not require synchronization?
%     \item How does the synchronization policy affect performance?
%     \item How do defects in the lattice affect the efficacy of \textit{Active} synchronization?
%     % \item Does \texttt{runahead} scheduling reduce the logical error rate?
% \end{itemize}

% We first describe the simulation methodology, after which we will present evaluations.

% \subsection{Growth of Synchronization Slack with Time}
% Applying non-Clifford gates on a logical qubit will desynchronize that logical qubit from the rest in a non-deterministic manner. Since CNOTs require Lattice Surgery, their use will spread the slack further. To quantify this spread of the slack between logical qubits, we used a Clifford+T decomposition~\cite{skd} of \textit{qft-12} and used a dataset of MWPM decoder latencies (which were normalized to be smaller than the code cycle time) to determine the slack introduced by every non-Clifford gates. Figure~\ref{fig:growth} shows how the mean slack normalized by the code cycle duration grows with time as more and more CNOTs are performed with the application of non-Clifford gates. Growth to the 0.5 mark shows that during the computation, about half the qubits will be out of sync. from others by almost one (or more) code cycles.

\subsection{\textit{Active} vs. \textit{Passive} Synchronization}

% \rev{
% \subsubsection{Color code Lattice Surgery}
% Since heterogeneous codes will cause desynchronization, we perform a Lattice Surgery operation between a color code patch ([[7, 1, 3]] Steane code) and a $d=3$ surface code patch, as shown in Figure~\ref{subfig:color-surface}~\cite{ito2024}. \textit{Active} synchronization reduces the overhead of synchronization by up to $\sim$7\%. This improvement will be higher for larger code distances, as will be evidenced in subsequent sections. The color code - surface code Lattice Surgery was decoded using the decoder presented in~\cite{lee2025color}.
% }

\subsubsection{LER reductions}
If the syndrome generation cycle times for patches $P$ and $P'$ are equal ($T_P = T_P'$), the \textit{Active} and \textit{Passive} policies are the only viable strategies to synchronize the two patches.
Figure~\ref{fig:sync_results} shows the reduction\footnote{\label{fn:improv}Reduction $=\frac{Logical Error Rate_{Passive}}{Logical Error Rate_{Active}}$.} in the logical error rate when \textit{Active} synchronization is used in place of \textit{Passive} before merging patches $P$ and $P'$ for IBM ((a) -- (b)) and Google-like((c) -- (d)) systems. \textbf{Since the worst-case slack ($\tau$) is bounded by the cycle time $\tau \leq T_{cycle}$}, we bounded $\tau \leq 1000ns$ to emulate superconducting systems since their cycle time $T_{cycle}$ is greater than 1000ns (Table~\ref{tab:configs}). Figure~\ref{fig:sync_results} shows that \textbf{the \textit{Active} policy achieves a reduction of up to 2.4x in the logical error rate}.

\begin{figure}[b]
    \centering
    \vspace{-0.1in}
    \includegraphics[width=\linewidth]{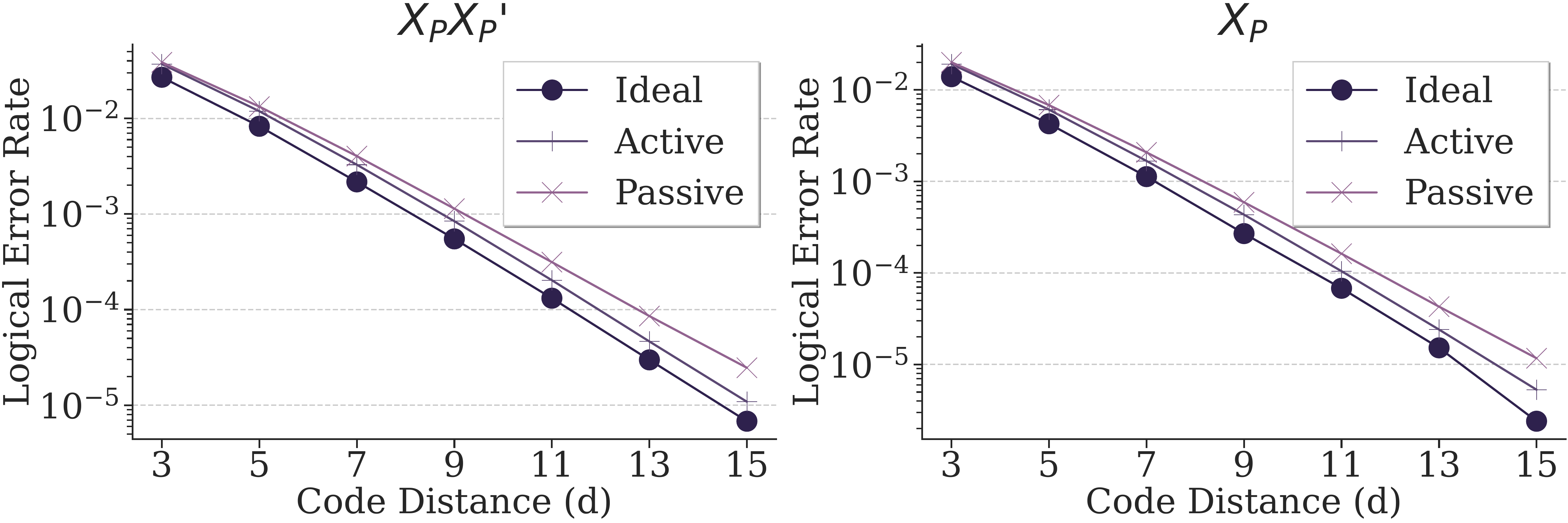}
    \caption{Logical error rate of the $X_{P}X_{P'}$ and $X_{P}$ observables with $Z$ basis Lattice Surgery after $2d+2$ rounds for an ideal system that does not require any synchronizations and a system either using \textit{Active} or \textit{Passive} synchronization. The worst-case synchronization slack of $\tau=$ 1000ns was chosen for this experiment with an IBM configuration ($p=10^{-3}$).}
    \Description[a figure]{}
    \label{fig:ler}
    
\end{figure}

\begin{figure}[t]
    \centering
    \includegraphics[width=\linewidth]{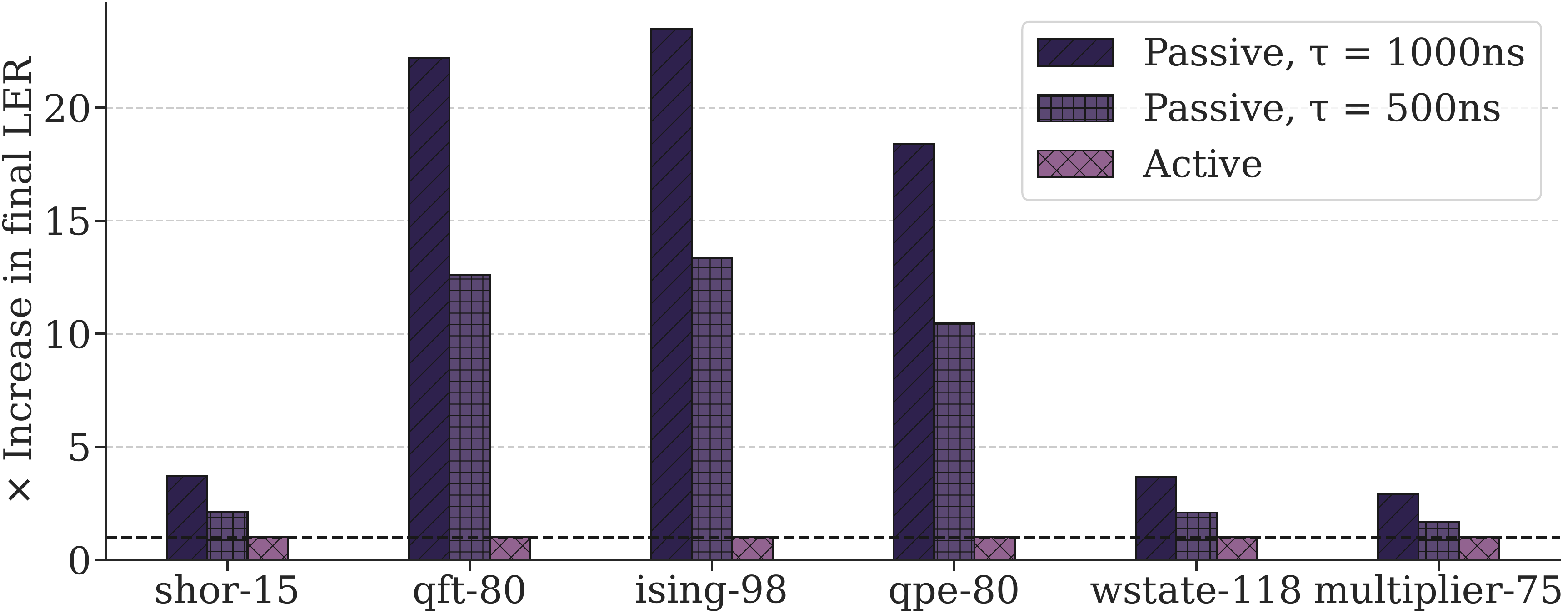}
    \caption{
    \rev{Relative increase in the final LER when the \textit{Passive} policy is used instead of \textit{Active} ($d=15$).}
    }
    \Description[a figure]{}
    \label{fig:bm_improvement}
    \vspace{-0.1in}
\end{figure}

\rev{
\subsubsection{Workload fidelity improvements}
The LER reduction reported in the previous section is for a single synchronization -- programs at scale will require hundreds of thousands or even millions of such synchronizations. To determine the \textbf{lower bound} on the LER improvement due to  \textit{Active} synchronization, we make a conservative assumption that error caused by idling during synchronization grows linearly\footnote{$(1 - \epsilon_{\text{l}})^{\#\text{ops}} \approx 1 - \#\text{ops} \cdot \epsilon_{\text{l}}$} with the number of Lattice Surgery operations. Even with this assumption, Figure~\ref{fig:bm_improvement} shows that the use of the \textit{Passive} policy can increase the final program LER by up to $~\sim$23$\times$, highlighting the necessity of the \textit{Active} policy.
}

\subsubsection{Cost of synchronization}
Compared to an ideal system in which all patches are always perfectly synchronized, what is the cost of synchronization in a more realistic system? Figure~\ref{fig:ler} shows the logical error rate for three systems: an ideal system that requires no synchronizations, and systems using the \textit{Passive} and \textit{Active} policies. For a worst-case synchronization slack of 1000ns, \textit{Active} synchronization can achieve a logical error rate that is much closer to the ideal system as compared to the \textit{Passive} policy, thus highlighting its effectiveness as a synchronization policy. 
    \emph{Despite not accounting for correlated and leakage errors caused by crosstalk and measurement~\cite{acharya2024quantumerrorcorrectionsurface, GoogleSuppressing}, there is a notable improvement in error suppression with Active synchronization. On actual hardware, the improvements with Active synchronization will likely be greater than reported. }
% \end{mybox}

\subsubsection{Comparison with the \textit{Active-intra} policy}
\begin{figure}[t]
    \centering
    % \hspace*{\fill}
    \begin{subfigure}[b]{0.49\linewidth}
        \centering
        \includegraphics[width=\textwidth]{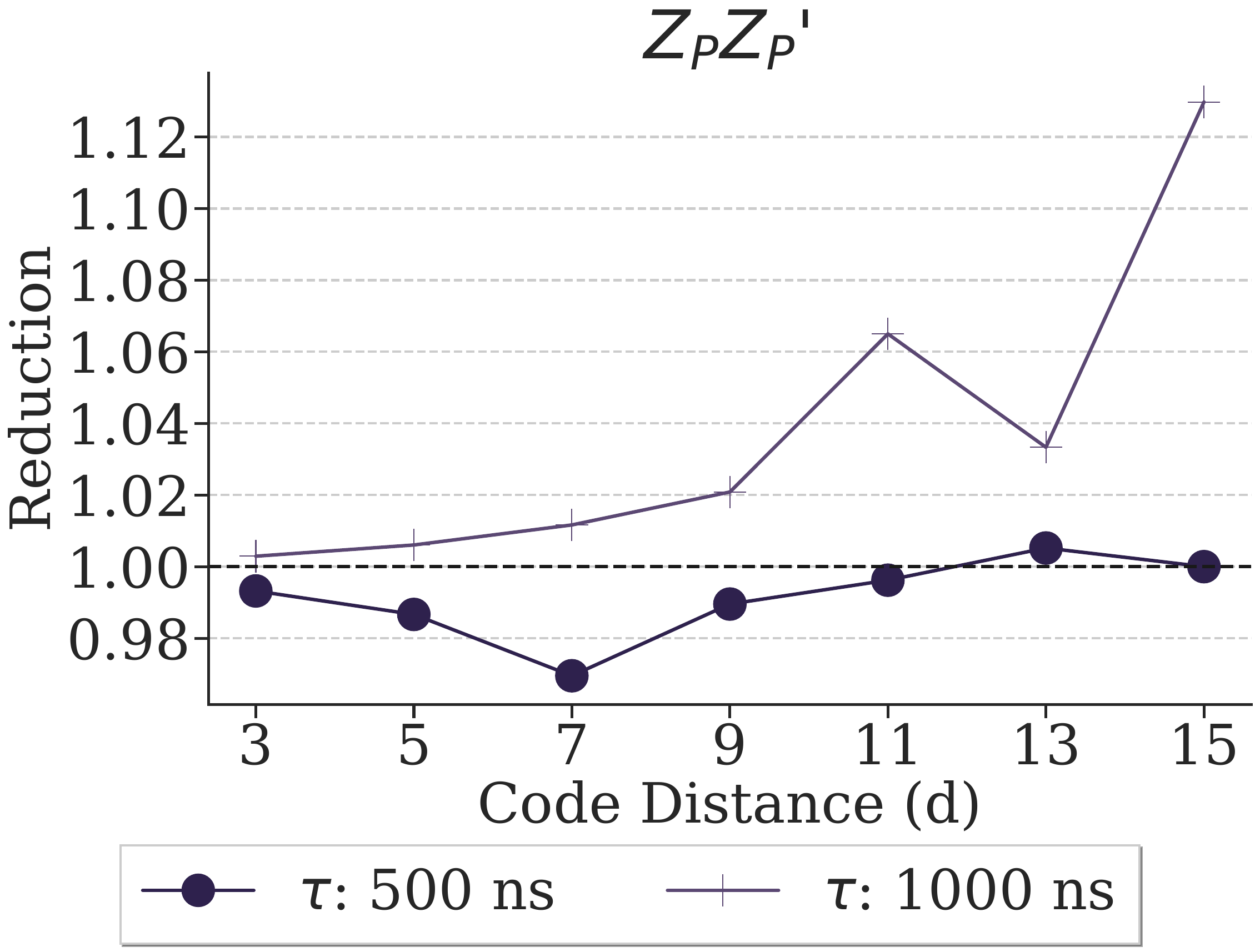}
        % \caption{}
        % \label{subfig:intra_x}
    \end{subfigure}
    \hspace*{\fill}
    \begin{subfigure}[b]{0.49\linewidth}
        \centering
        \includegraphics[width=\textwidth]{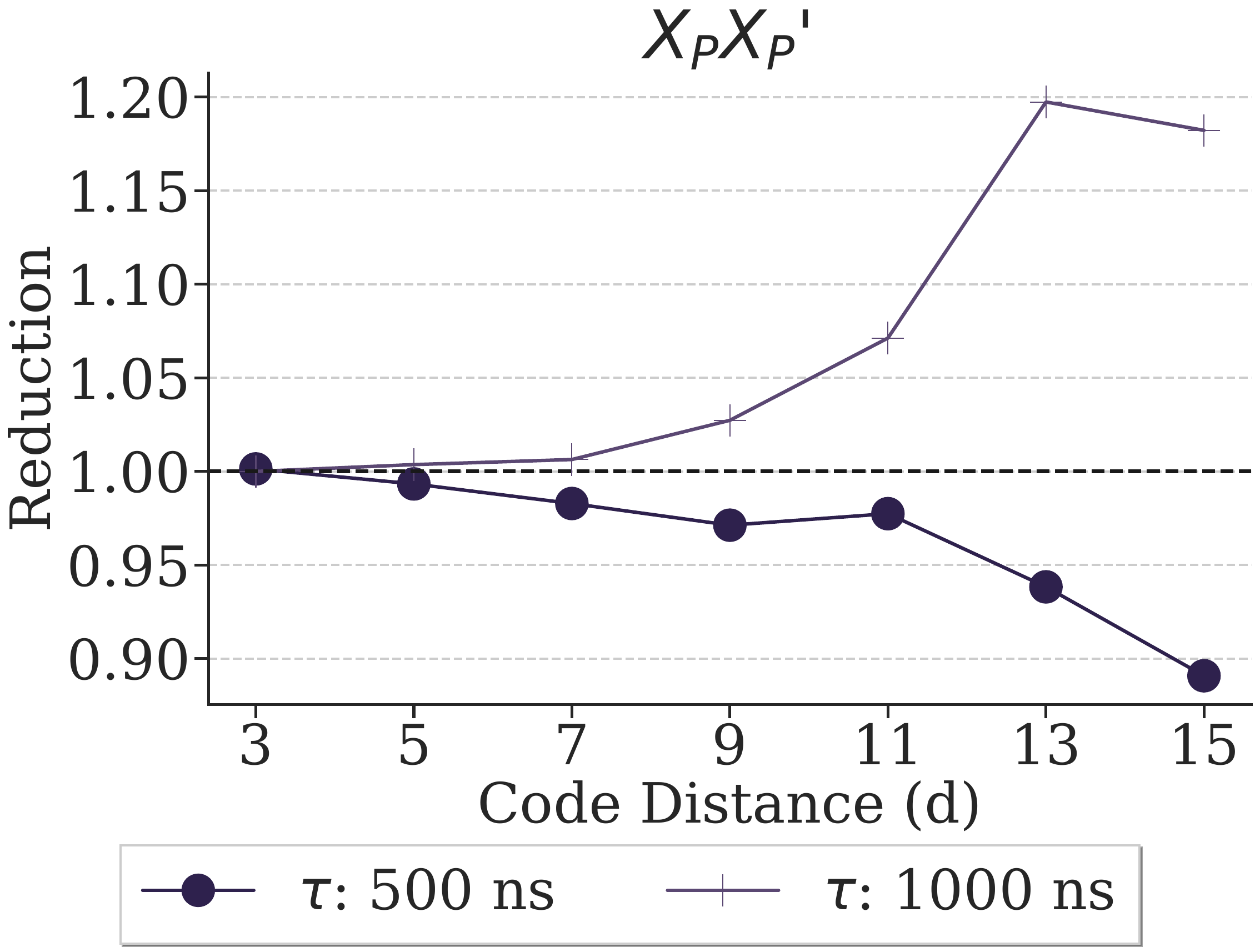}
        % \caption{}
        % \label{subfig:intra_z}
    \end{subfigure}
    % \hspace*{\fill}
    \caption{
    Reduction in the LER with the \textit{Active-intra} policy for $X$ and $Z$ basis Lattice Surgery. The \textit{Active-intra} policy leads to an \textbf{increase} in the LER for some cases.
    }
    \vspace{-0.15in}
    \Description[intra]{}
    \label{fig:result_intra}
\end{figure}

The \textit{Active-intra} policy distributes the synchronization slack within the \textbf{last round} before Lattice Surgery. As shown in Figure~\ref{fig:result_intra}, this can either exacerbate the LER or lead to very tiny reductions compared to the \textit{Active} policy (IBM configuration). This is because the \textit{Active-intra} policy also introduces decoherence errors on all measure qubits, which increases the chances of logical errors.

\subsubsection{Will running additional rounds help the \textit{Active} policy?}
\begin{figure}[t]
    \centering
    \begin{subfigure}[b]{0.46\linewidth}
        \centering
        \includegraphics[width=\textwidth]{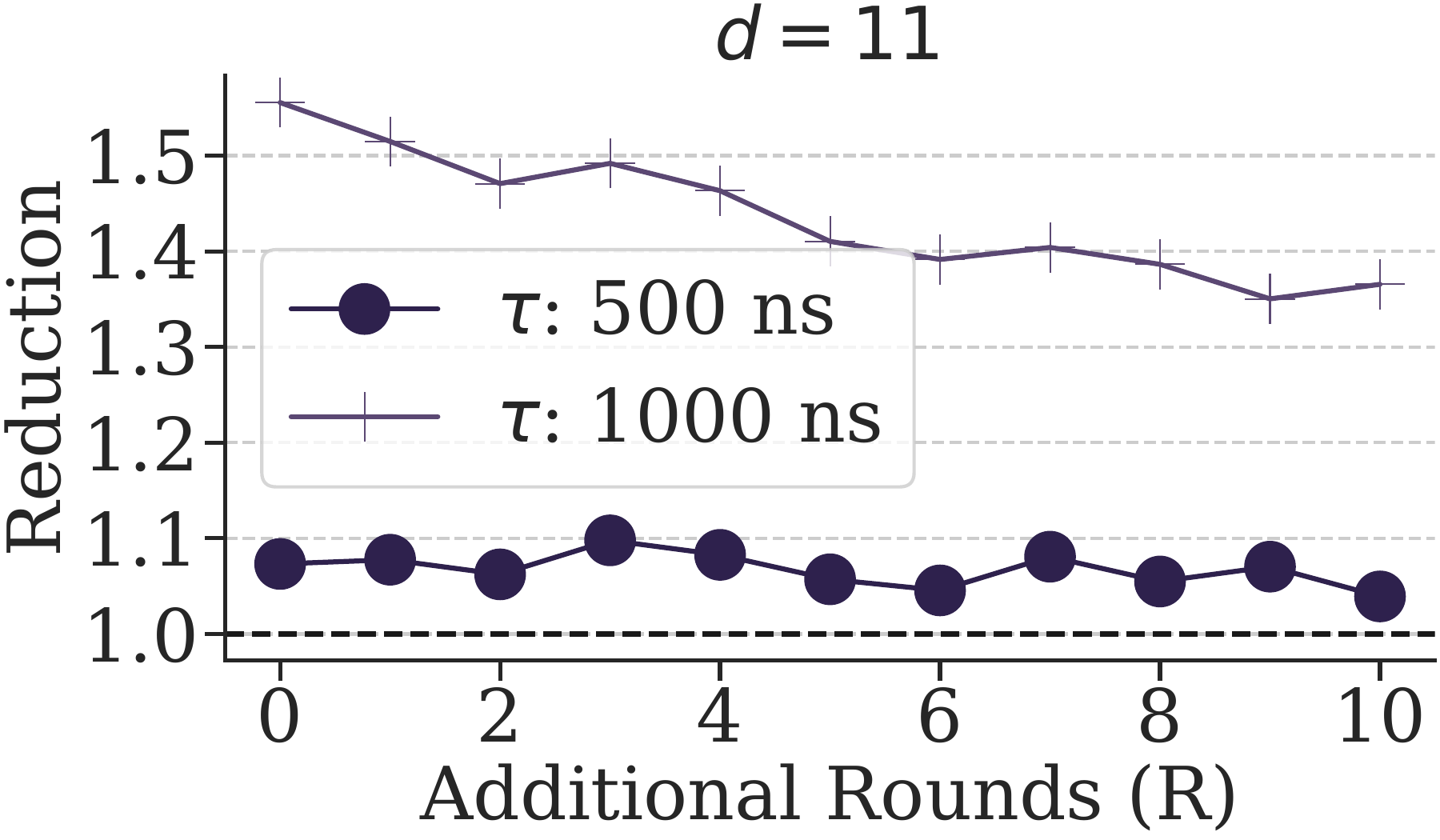}
        \caption{}
        \label{subfig:active_additional_rounds}
    \end{subfigure}
    \hspace*{\fill}
    \begin{subfigure}[b]{0.46\linewidth}
        \centering
        \includegraphics[width=\textwidth]{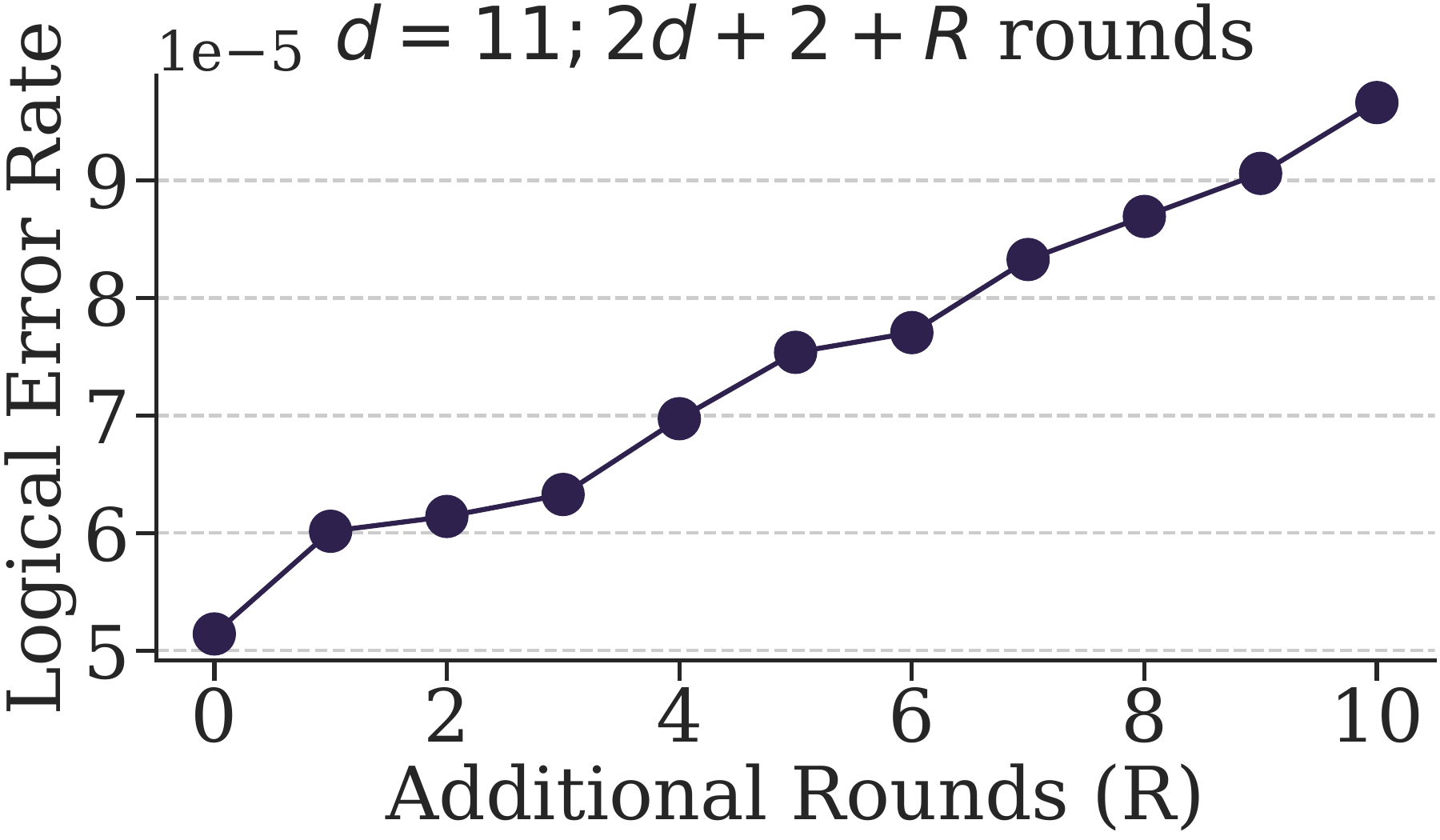}
        \caption{}
        \label{subfig:ler_rounds}
    \end{subfigure}
    \caption{
    (a) Reduction in the LER compared to the \textit{Passive} policy when the \textit{Active} policy distributes its slack over $d + 1 + R$ rounds rather than $d+1$ rounds;
    (b) Influence of the number of rounds run on the LER (without any slack/idling).
    }
    \vspace{-0.1in}
    \Description[a figure]{}
    \label{fig:active_additional_rounds}
\end{figure}

\textit{Active} synchronization works by dividing the slack into smaller idle periods and slowing the leading logical patch gradually. If the number of rounds over which the slack is distributed is increased, the idle periods before every round become even smaller. To evaluate the effect of additional rounds on the efficacy of \textit{Active} synchronization, we added $R$ additional rounds before Lattice Surgery was performed, thus allowing the slack to be distributed over $d + 1 + R$ rounds for the \textit{Active} policy. The \textit{Passive} policy was also used after $d + 1 + R$ rounds. This setup corresponds to starting the synchronization process \textit{earlier} in time, since more rounds are being used to distribute the slack for the \textit{Active} policy. Figure~\ref{subfig:active_additional_rounds} shows how the benefit of \textit{Active} synchronization diminishes as additional rounds are run, especially for $\tau=1000ns$ (averaged over both $X_{P}X_{P'}, Z_{P}Z_{P'}$ observables, with the IBM configuration). From Figure~\ref{fig:active_additional_rounds}, we can establish that increasing the number of rounds over which the \textit{Active} policy operates has diminishing returns. This can be understood from Figure~\ref{subfig:ler_rounds}, where the LER increases with the number of additional rounds run, even without the presence of any synchronization slack and idling. This is because decoders are imperfect, and these imperfections lead to error accumulating after every round.

\subsection{The \textit{Extra Rounds} and \textit{Hybrid} Policies}
\subsubsection{LER reduction}
\begin{figure}[t]
    \centering
    \includegraphics[width=0.8\linewidth]{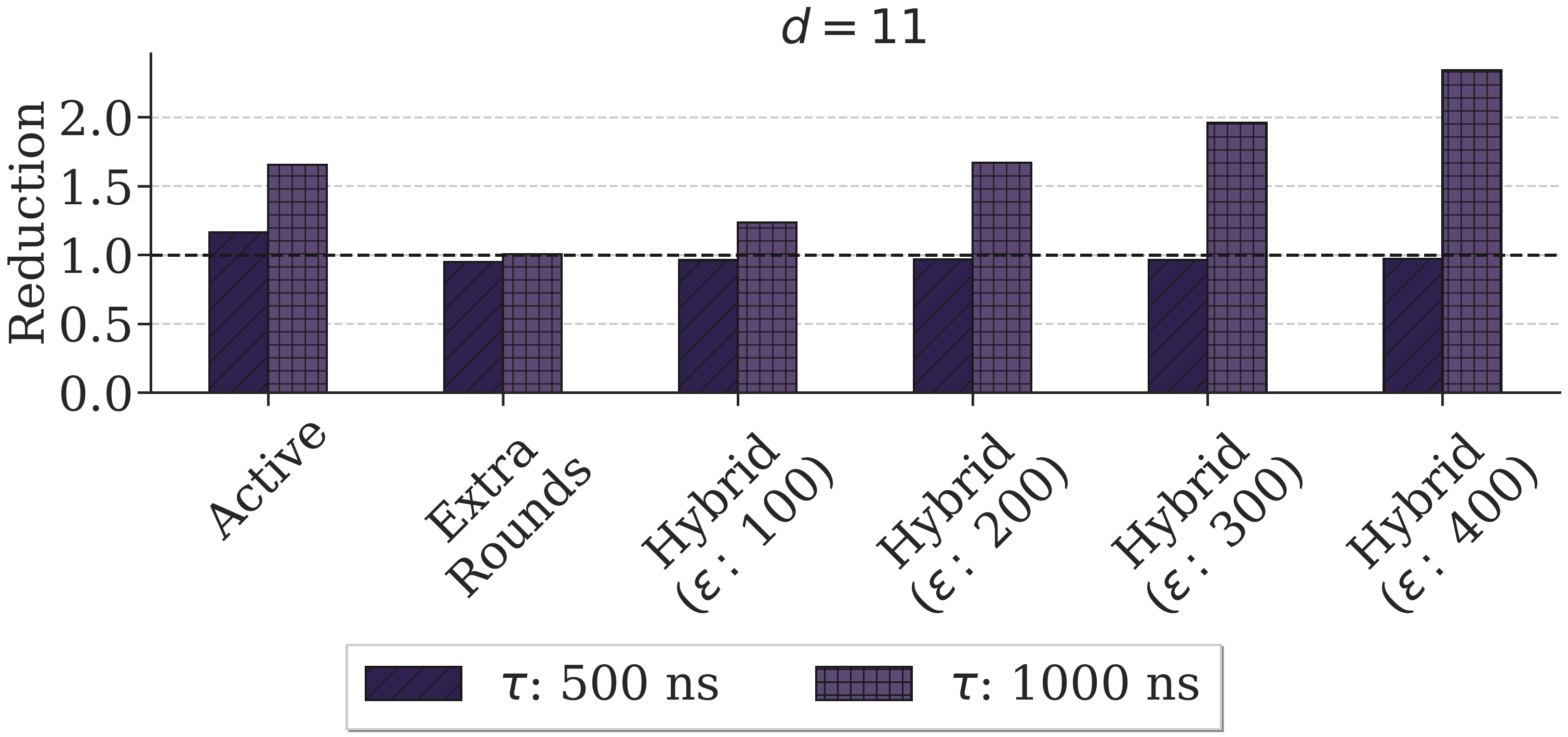}
    \caption{
    \rev{Reduction in the LER compared to the \textit{Passive} policy for different initial slacks $\tau$ and tolerances $\epsilon$ (in ns). The \textit{Hybrid} policy works more effectively for larger $\tau$.}
    }
    \vspace{-0.1in}
    \Description[a figure]{}
    \label{fig:hybrid_results}
\end{figure}

So far, we have shown how the \textit{Active} policy is superior to the \textit{Passive} policy, especially for large $\tau$. If the syndrome generation cycle times $T_P, T_P'$ of the two patches are different, additional policies are possible, as described in Section~\ref{sec:sync_two}. Figure~\ref{fig:hybrid_results} shows the reduction in the LER for different policies. We varied the slack tolerance for the \textit{Hybrid} policy from 100 -- 400ns and used all combinations of the cycle times -- $T_P=1000ns, T_P'=1050/1100/1150ns$. These cycle times are representative of cycle times consisting of 1/2/3 additional CNOTs compared to the surface code, as would be the case for color/qLDPC codes. The results were obtained as an average over all combinations of cycle times and the $X_{P}X_{P'}, Z_{P}Z_{P'}$ observables (IBM configuration). Figure~\ref{fig:hybrid_results} shows that for  larger values of $\tau$, the \textit{Hybrid} policy works much better, since it allows the slack to be reduced significantly. Furthermore, higher values of $\epsilon$ allow more leeway in selecting the number of extra rounds to be run, allowing significantly better LERs. Conversely, for smaller values of $\tau$, Figure~\ref{fig:hybrid_results} suggests that the \textit{Active}/\textit{Passive} policies would be more advantageous. Table~\ref{tab:hybrid_result} shows that the reduction is even higher at 3.4$\times$ for $d=15$.

\begin{table}[t]
\centering
\caption{Average reduction in the LER for all policies compared to the \textit{Passive} policy for $d=11,13,15$.}
\label{tab:hybrid_result}
\resizebox{0.9\linewidth}{!}{%
\begin{tabular}{c|cccc}
\specialrule{1.5pt}{1pt}{1pt}
\multirow{4}{*}{\begin{tabular}[c]{@{}c@{}}\textbf{LER Reduction}\\ ($\tau=1000ns$)\end{tabular}} & $d$                    & \textit{Active} & \textit{Extra Rounds} & \cellcolor{green!25}\textit{Hybrid ($\epsilon=400ns$)} \\ \cline{2-5} 
                                                                                                &  $d=15$                & 2.14            & 1.63                  & \cellcolor{green!25}3.4    \\         
                                                                                                &  $d=13$                & 1.87            & 1.29                  & \cellcolor{green!25}2.72    \\ 
                                                                                                &  $d=11$                & 1.65            & 1.07                  & \cellcolor{green!25}2.34    \\\specialrule{1.5pt}{1pt}{1pt}
\end{tabular}
\vspace{-0.25in}
}
\end{table}

\begin{figure}[t]
    \centering
    \includegraphics[width=0.9\linewidth]{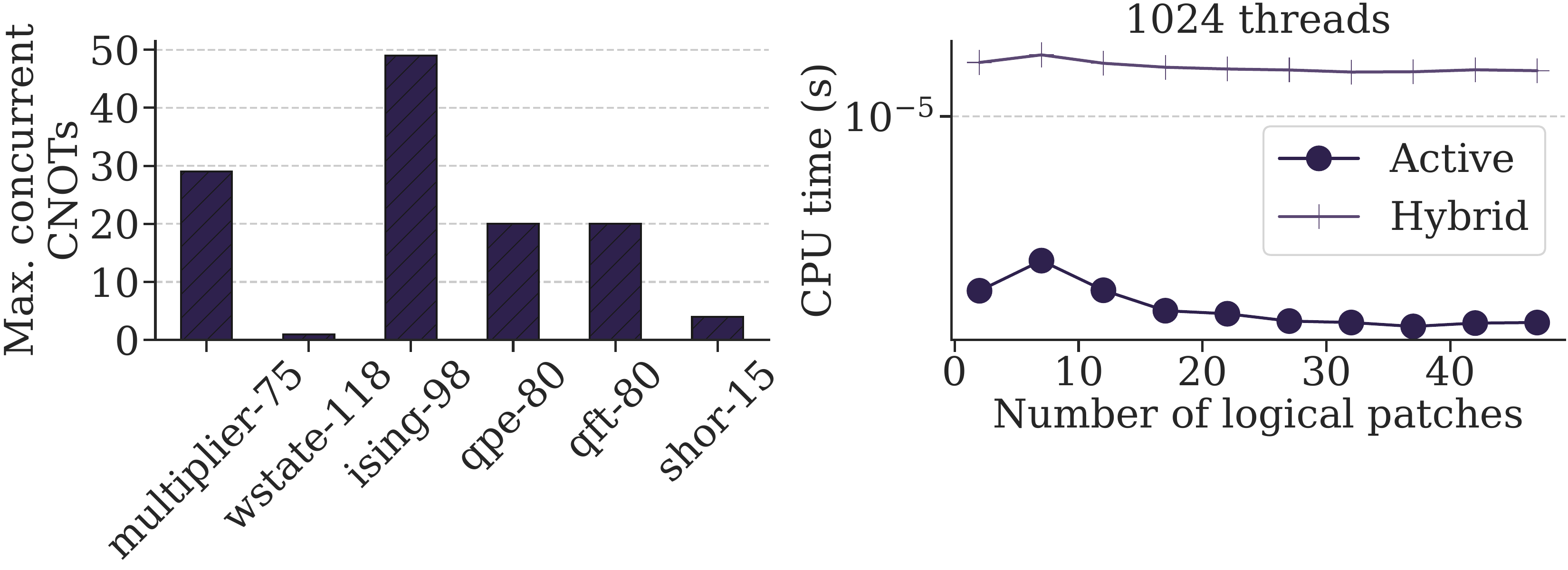}
    \caption{
    \rev{Time required to synchronize all patches as a function of the number of patches in the system.}
    }
    \vspace{-0.1in}
    \Description[a figure]{}
    \label{fig:compile_time}
\end{figure}

\rev{
\subsubsection{Compilation time for synchronization}
\label{sec:compilation_time}
How long will it take to determine the slack and the number of additional rounds required by the \textit{Hybrid} policy? We run a numerical simulation in software and determine the average time it takes to compute the slack and additional rounds for up to 50 logical patches, as shown in Figure~\ref{fig:compile_time}. We set the upper bound to 50 from Figure~\ref{fig:compile_time} which shows the maximum number of concurrent CNOTs (and hence synchronized Lattice Surgery operations). As discussed in Section~\ref{sec:syncK}, all patches that require synchronization can be synchronized pair-wise in parallel, which results in a \textbf{constant time ($O(1)$)} overhead. 
% However, the number of patches that can be synchronized concurrently will depend on the hardware available, and this limit will likely result in a quadratic time ($O(\binom{k}{2}/n)$) overhead (where $k$ is the number of patches and $n$ is the number of processors).
}

\subsection{Synchronization in Neutral Atom Platforms}
\begin{figure}[t]
    \centering
    \includegraphics[width=0.9\linewidth]{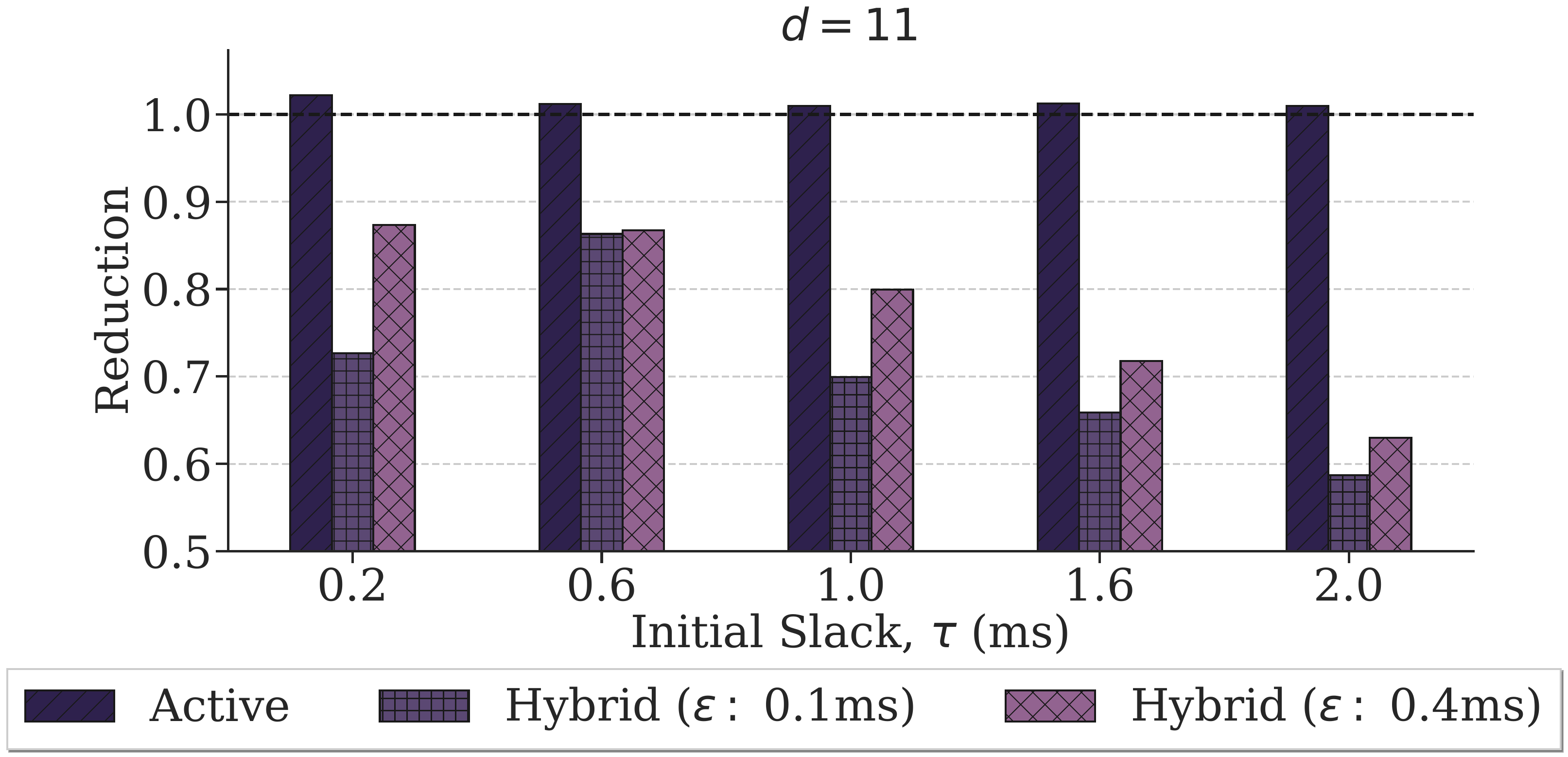}
    \caption{Reduction in the LER compared to the \textit{Passive} policy with the use of the \textit{Active} and \textit{Hybrid} policies on a neutral atom system. Running additional rounds, even with the \textit{Hybrid} policy, is detrimental to the LER.}
    \vspace{-0.1in}
    \Description[a figure]{}
    \label{fig:na_sync}
\end{figure}

So far, we have discussed results on superconducting systems. Neutral atom systems have considerably higher $T_1, T_2$ times (Table~\ref{tab:configs}) which reduces the impact of idling errors. Figure~\ref{fig:na_sync} shows the reduction in the LER of all policies for different initial synchronization slacks compared to the \textit{Passive} policy\footnote{The results were averaged over the $X_{P}X_{P'}, Z_{P}Z_{P'}$ observables and all combinations of $T_P=2ms$, $T_P'=2.2ms, 2.4ms, 2.6ms$.}. \textit{Active} synchronization achieves a marginal $\sim2\%$ reduction in the LER over the \textit{Passive} policy, but the \textit{Hybrid} policy leads to an increase in the LER due to more rounds being run, as shown in Table~\ref{tab:na_rounds}. These results indicate that the \textit{Active}/\textit{Passive} policies are always better on neutral atom systems due to their long coherence times.

\begin{table}[ht!]
\centering
\caption{Additional rounds needed for synchronization with the \textit{Hybrid} policy -- these result in an increase in the LER.}
\label{tab:na_rounds}
\resizebox{0.9\linewidth}{!}{%
\begin{tabular}{l|ccccc}
\specialrule{1.5pt}{1pt}{1pt}
\textbf{Initial Slack $\tau$ (ms)}             & \textbf{$\tau=0.2$} & \textbf{$\tau=0.6$} & \textbf{$\tau=1.0$} & \textbf{$\tau=1.6$} & \textbf{$\tau=2.0$} \\ \hline
\textit{\textbf{Hybrid ($\epsilon=0.1ms$)}} & 9                   & 3                   & 6                   & 8                   & 12                  \\
\textit{\textbf{Hybrid ($\epsilon=0.4ms$)}} & 5                   & 3                   & 5                   & 8                   & 10                  \\ \specialrule{1.5pt}{1pt}{1pt}
\end{tabular}
    \vspace{-0.1in}
}
\end{table}

% \begin{mybox}
% % \centering
%     \emph{For both superconducting and neutral atom technologies, simply running extra rounds of error correction for synchronization may be suboptimal. This is because (i) every additional round accumulates errors and degrades the LER and (ii) running additional rounds slows down the computation (especially in neutral atom systems which have slow operational latencies).}
% \end{mybox}

\subsection{Performance Improvements}
\begin{figure}[t]
    \centering
    \includegraphics[width=0.9\linewidth]{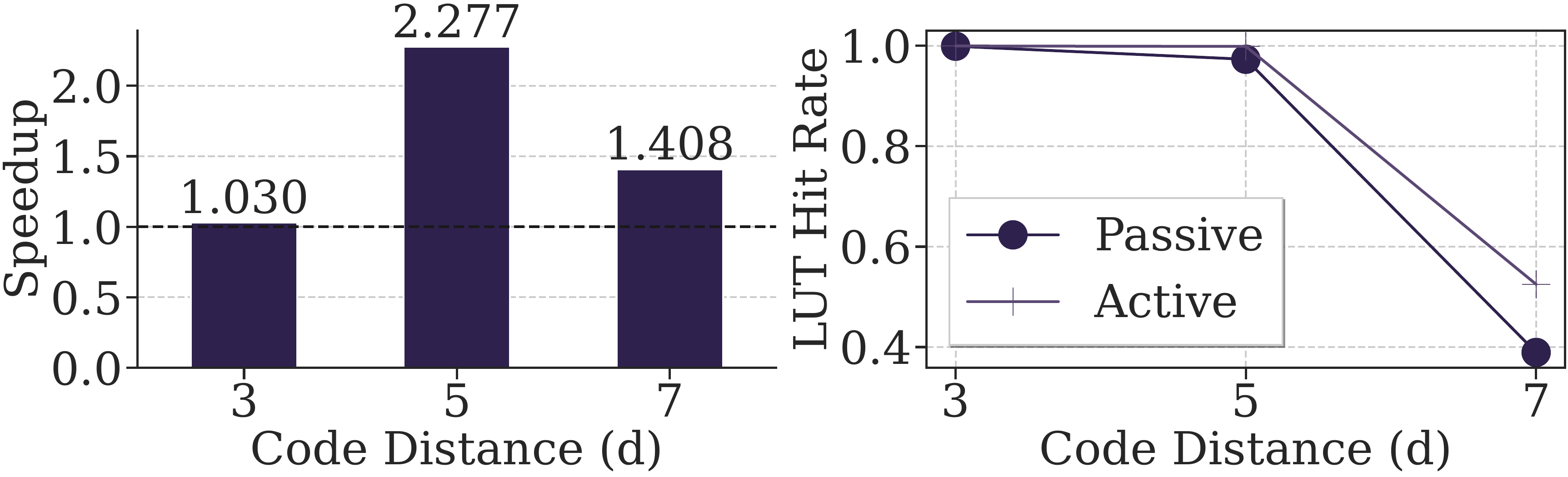}
    \caption{Speed-up gained by \textit{Active} synchronization over \textit{Passive} per Lattice Surgery operation and the corresponding LUT hit rates.}
    \Description[a figure]{}
    \label{fig:speedup}
    \vspace{-0.25in}
\end{figure}

\textit{Active} synchronization outperforms \textit{Passive} synchronization in terms of the logical error rate, but how does this impact the system performance? To evaluate its benefits, we considered a hierarchical decoder~\cite{delfosse2020hierarchical} where the fast, low overhead decoder was a Lookup Table (LUT) based decoder~\cite{lilliput, Svore2014} and the slow, accurate decoder was the MWPM decoder. If a syndrome missed the LUT, the MWPM decoder was invoked, thus requiring more time for the error to be decoded. As a LUT of infinite size is not practical, we limited the size to 3KB, 3MB, and 30MB for $d=3,5,7$ respectively (beyond $d=7$, the LUT size required is impractical~\cite{lilliput}). Figure~\ref{fig:speedup} shows the relative speedup of \textit{Active} synchronization over \textit{Passive} synchronization due to the reduction in error rate (thus hard-to-decode errors). This speedup was determined numerically over 1B trials with the assumption that a hit in the LUT incurs a decoding latency of 20ns while a miss invokes the MWPM decoder, whose latency is sampled from a MWPM latency dataset. For $d=3$, the LUT was able to capture almost all possible syndromes for both cases, thus yielding a small speedup of 0.3\%. However, for $d=5$, the speedup is more than 2.2$\times$ as more syndromes hit in the LUT in the case of \textit{Active} synchronization. For $d=7$, the hit rate of the LUT is significantly lower, reducing the overall speedup offered by \textit{Active} synchronization, as shown in Figure~\ref{fig:speedup}. 

\section{Related Work}

This work is one of the first to explore the need and means of synchronization for FTQC systems. 
The closest related work is by Xuereb et al., who examined the effects of imperfect timekeeping on quantum circuits~\cite{Xuereb2023}. 
{Silveri et al.~\cite{Silveri2022} discuss the {fundamental problem of protecting qubits from idling errors}.} 
% while ensuring fast, responsive gates, further highlighting how idling errors are problematic.}
The following works are broadly connected to ours.

% on architectures and compilers for fault-tolerant systems are orthogonally related to proposed synchronization and scheduling policies. 

\subheading{Decoder architectures}
Decoders need to be accurate and fast for practical fault-tolerant quantum computing using the surface code. Neural-network based~\cite{ueno2022neoqec, varbanov2023neural, nndecoder, zhang2023, googleRNN} and hierarchical decoders~\cite{delfosse2020hierarchical, caune2023belief, chamberland2022techniques, smith2023predecoder} accelerate decoding with reasonable hardware requirements. AFS~\cite{afs}, QECOOL~\cite{Ueno2021}, and Clique~\cite{clique} are other decoder architectures that explore using SFQ logic. Qualatis~\cite{qualitis} used SFQ logic to design a decoder supporting Lattice Surgery. Other low-latency decoders have also been proposed~\cite{lilliput, astrea, Alavisamani2024}.

\subheading{System and compiler architectures}
XQSim~\cite{xqsim} is a full-system simulation model for an SFQ-based controller for fault-tolerant quantum computing. Prior works study mapping and scheduling stabilizer circuits~\cite{Wu2022}, and routing logical operations~\cite{hua2021autobraid,javadi2017optimized}, and focused on system architecture for FTQC~\cite{ding2018magic,stein2023multi,stein2023microarchitectures,choi2023fault,duckering2020virtualized,kim2024}.

\section{Conclusions}

In this paper, we introduce the problem of synchronization in FTQC systems stemming from the use of heterogeneous codes, qubit/coupler dropouts, and techniques such as twist-based Lattice Surgery. Lattice Surgery operations between two or more logical qubits require the syndrome generation cycles of all logical qubits involved to be synchronized -- this necessitates synchronization policies that can perform synchronization at runtime. Synchronization of two or more logical qubits will require the leading logical qubit to pause/slow down its syndrome generation cycles to allow the lagging logical qubit(s) to catch up. We call the simplest policy the \textit{Passive} policy, where the leading logical qubit waits for the lagging logical qubit right before Lattice Surgery. To reduce the detrimental impact of idling errors, we propose an \textit{Active} synchronization policy that slows the leading logical qubit \textit{gradually} by splitting the slack into smaller chunks and distributing it between multiple code cycles. This policy achieves a reduction of up to 2.4$\times$ in the logical error rate. We augment this policy by running a few additional rounds of error correction, which we call the \textit{Hybrid} policy, yielding a reduction of up to 3.4$\times$ in the logical error rate. Furthermore, reducing the impact of idling errors helps the decoding latency, resulting in a decoding speedup of up to 2.2$\times$.

\begin{acks}
    The authors would like to thank Benjamin Lienhard and Michael Beverland for their feedback on previous versions of this paper. The authors would also like to thank the late Burton Smith, whose encouragement motivated this exploration. The authors also thank the Center for High Throughput Computing (CHTC) at the University of Wisconsin-Madison and Guri Sohi for their help with computing infrastructure. This research was funded by an NSF Career Award \#2340267.
\end{acks}

% Finally, by optimizing the gate schedule within a logical qubit with \texttt{runahead} scheduling, the effectiveness of \textit{Active} synchronization is improved by up to 20\%. ()

% The first part of our work focuses on lowering idling within the surface code cycle mapped on hardware with non-uniform gate latencies. We propose \texttt{runahead} scheduling that executes operations in an as-soon-as-possible manner. 
% Compared to the default lockstep execution model, our scheduler can reduce the idling by up to 60\% and reduce logical error by 80\%. 
% The second part our work focuses on the idling induced during logical operations, which require patches to have synchronized QEC cycles. We present an \textit{Active} synchronization policy to preemptively synchronize logical qubits by distributing idling across QEC cycles to reduce its impact.  

\bibliographystyle{ACM-Reference-Format}
\bibliography{references}

\appendix
\section{Artifact Details}

\subsection{Summary}
This artifact contains the code to verify the key results of this paper. Specifically, this artifact will generate and plot data for Figures~\ref{fig:cult_slack}, \ref{fig:sync_results}, \ref{fig:ler}, \ref{subfig:active_additional_rounds}, \ref{fig:hybrid_results}, and \ref{fig:na_sync}. Additionally, the artifact contains code to verify the results of Figure~\ref{fig:ibm_Active_Passive_exp}. Note that this will require access to IBM quantum systems, and all results might be impossible to obtain with just the open access plan. Finally, the artifact also contains a Python notebook \texttt{synchronization.ipynb} that can be used with the Azure Quantum Resource Estimation toolbox to obtain results that correspond to Figure~\ref{subfig:motive-4}. 

All details, including hardware requirements and runtime, are summarized below:

\begin{itemize}
    \item \textbf{License}: MIT License
    \item \textbf{Software requirements}: Linux system, Docker, Python
    \item \textbf{Hardware requirements}: Multi-core CPU $\ge 128$ cores
    \item \textbf{Memory requirements}: At least 512 GB of RAM
    \item \textbf{Disk space requirements}: Not more than 5GB
    \item \textbf{Runtime environment}: Docker container, Python scripts
    \item \textbf{Time needed}: At least 5 days with 128 cores. Will require more time with fewer cores
    \item \textbf{Publicly available}: Yes
    \item \textbf{Artifact DOI}: 10.5281/zenodo.15092177
\end{itemize}

\subsection{Setting up the artifact}
Download the artifact in your system from \url{https://doi.org/10.5281/zenodo.15092177}. Extract the tarball, and then build the Docker container and start an interactive session using the following commands: 

% \begin{verbatim}
% $    docker build -t $USER/lattice-sim .
% $    docker-compose run app
% \end{verbatim}
\begin{lstlisting}
$    docker build -t $USER/lattice-sim .
$    docker-compose run app
\end{lstlisting}

Since these are long-running experiments, we recommend running the container and the experiments within a \texttt{tmux} session.

\subsection{Instructions for generating main results}

Each Python script in the artifact takes two command line inputs: (i) the total number of shots that need to be run for the simulation, and (ii) the number of processes to spawn to complete all simulations. Internally, each script divides the total number of shots into smaller batches, and runs a batch using the specified number of processors for an appropriate number of iterations. Every Python script can be run using the template below ({\color{red} \textbf{All scripts must be run within the Docker container}}): 

\begin{lstlisting}
$   python3 <script>.py <# shots> <# processes>
\end{lstlisting}

To simplify the data generation process, we have added a bash script \texttt{run\_all.sh} that will run all Python scripts with a recommended number of shots and 128 processes. To run this script, simply run:

\begin{lstlisting}
$   bash run_all.sh
\end{lstlisting}

\subsection{Plots}
Once all Python scripts have been executed, all plots to verify key results can be created using the Python notebook \texttt{plots.ipynb}. This notebook is self-sufficient and can be run outside the Docker container (it will install all required Python dependencies into your local Python environment). 

{\color{red} \textbf{Note}}: Results are very sensitive to the number of shots experiments are run for (especially for larger code distances $d=13, 15$). Ideally, the results of Figure~\ref{fig:sync_results} should be plotted after running at least 100M shots, but time constraints could prevent this from being possible. Thus, expect to see some deviations for larger code distances for a smaller number of shots.

\subsection{Optional results}
\texttt{plots.ipynb} also contains code to verify the results of Figure~\ref{fig:ibm_Active_Passive_exp}. Running this code will require access to IBM quantum systems, and will most likely need more than the 10-minute quota of the free open-access plan. We have added the code in case reviewers are interested in running the experiments, and all Qiskit dependencies can also be installed through the notebook. 

Figures~\ref{subfig:motive-4} and \ref{fig:bm_improvement} are derived from results obtained from the Azure Quantum Resource Estimator (\url{https://learn.microsoft.com/en-us/azure/quantum/intro-to-resource-estimation}). We have provided the Python notebook \texttt{synchronization.ipynb} that can be uploaded to the Azure interface to obtain the results.

\subsection{Methodology}
Submission, reviewing, and badging methodology: 
\begin{itemize}
    \item \url{http://ctuning.org/ae/submission-20201122.html}
    \item \url{http://ctuning.org/ae/reviewing-20201122.html}
    \item \url{https://www.acm.org/publications/policies/artifact-review-badging}
\end{itemize}
\end{document}